\documentclass[a4paper,11pt]{article}

\usepackage{amsmath}
\usepackage{amssymb}
\usepackage{mathtools}

\usepackage[linktocpage]{hyperref}
\usepackage[dvipsnames,table]{xcolor}
\usepackage{cancel,ulem}
\hypersetup{
    colorlinks,
    linkcolor={Mulberry},
    citecolor={Mulberry},
    urlcolor={Mulberry}
}

\def\gtwid{\mathrel{\raise.3ex\hbox{$>$\kern-.75em\lower1ex\hbox{$\sim$}}}}
\def\ltwid{\mathrel{\raise.3ex\hbox{$<$\kern-.75em\lower1ex\hbox{$\sim$}}}}
\def\square{\kern1pt\vbox{\hrule height 1.2pt\hbox{\vrule width 1.2pt\hskip 3pt
   \vbox{\vskip 6pt}\hskip 3pt\vrule width 0.6pt}\hrule height 0.6pt}\kern1pt}
   
\newcommand{\scr}{\scriptscriptstyle}
\newcommand{\longsim}{\scalebox{1.8}[1]{$\sim$}}

\usepackage[margin=2.5cm]{geometry}

\numberwithin{equation}{section}

\usepackage{graphicx}
\graphicspath{ {./figures/} }

\allowdisplaybreaks

\usepackage{cite}

\usepackage[labelfont=bf]{caption}

\usepackage{orcidlink}

\begin{document}

\hfill {\small UFIFT-QG-26-07}

\smallskip

\begin{center}

\renewcommand{\thefootnote}{\fnsymbol{footnote}}

{\Large \bf 
Massive scalar's self-mass from gravitons in de Sitter
}

\setcounter{footnote}{0} 

\bigskip

Dra\v{z}en Glavan\,\orcidlink{0000-0002-1983-0448},${}^{a,b,}$\footnote[1]{email: 
	\href{mailto:glavan@fzu.cz}{\tt d.glavan@uu.nl}}
Shun-Pei Miao\,\orcidlink{0000-0002-0754-6888},${}^{c,}$\footnote[2]{email:
	\href{mailto:spmiao5@mail.ncku.edu.tw}{\tt spmiao5@mail.ncku.edu.tw}}
Tomislav Prokopec\,\orcidlink{0000-0003-0391-5743},${}^{b,}$\footnote[3]{email: 
	\href{mailto:t.prokopec@uu.nl}{\tt t.prokopec@uu.nl}}
Richard P.~Woodard\,\orcidlink{0000-0003-0830-1396}${}^{\,d,}$\footnote[4]{email: 
	\href{mailto:woodard@phys.ufl.edu}{\tt woodard@phys.ufl.edu}}

\bigskip

{\it 
${}^a$\,CEICO, Institute of Physics of the Czech Academy of Sciences (FZU),
\\
Na Slovance 1992/2, 182 21 Prague 8, Czech Republic

\smallskip

${}^b$\,Institute for Theoretical Physics, Spinoza Institute \& EMME$\Phi$,
Utrecht University,
\\
Postbus 80.195, 3508 TD Utrecht, The Netherlands

\smallskip

${}^c$\,Department of Physics, National Cheng Kung University,
\\
No.~1 University Road, Tainan City, 70101, Taiwan

\smallskip

${}^d$\,Department of Physics, University of Florida, Gainesville, FL 32611, U.S.A.
}

\

\medskip

\parbox{0.95\linewidth}{
We compute the fully renormalized one-graviton-loop contribution to the self-mass of 
a massive scalar field in de Sitter space. The computation is performed using dimensional 
regularization and the divergences are absorbed by BPHZ counterterms. This self-mass can 
be used to study one-loop quantum-gravitational corrections to the dynamics of massive
spectator scalars in inflation.
}

\end{center}

\medskip

\hrule
\tableofcontents

\pagebreak

\section{Introduction}

Gravitational particle production
\cite{Parker:1968mv} during inflation generates a vast
ensemble of long-wavelength inflationary gravitons
\cite{Starobinsky:1979ty}. This enhances
quantum-gravitational corrections to matter fields
propagating during inflation, although they remain small in
absolute terms, with a typical magnitude estimated
 to be $G_{\rm \scr N}H^2 \!\sim\! 10^{-11}$.
Explicit computations have nevertheless revealed an
additional enhancement in the form of large temporal or
spatial logarithms
\cite{Miao:2006gj,Glavan:2013jca,Wang:2014tza,
Tan:2021lza,Glavan:2021adm,Tan:2022xpn,Glavan:2016bvp}. 
In all cases
the calculations concerned one-graviton-loop corrections to
massless fields, which motivates extending the investigation
to massive fields.

Massive scalar fields in de Sitter space exhibit two
qualitatively distinct regimes, distinguished by the scalar
mass relative to the Hubble rate. Light massive scalars
retain many infrared features of the massless minimally
coupled case, and their quantum-gravitational corrections may
therefore exhibit large enhancements as well. The precise
form of these enhancements may differ from the massless
case, as illustrated by the self-interacting light scalar
\cite{Kamenshchik:2021tjh}, and can only be determined by
explicit computations, which we leave for future work.

Our immediate interest lies in the heavy-field regime.
Heavy fields during inflation have been studied extensively
within the cosmological collider program
\cite{Arkani-Hamed:2015bza}. Our motivation is different:
we use massive scalar fields to model sources and observers
in the construction of gauge-independent
quantum-gravitational corrections to exchange potentials in
de Sitter space, in analogy with massive source and observer
fields in flat-space S-matrix calculations.

Most of the logarithm-enhanced corrections thus far \cite{Miao:2006gj,Glavan:2013jca,Wang:2014tza,
Tan:2021lza,Glavan:2021adm,Tan:2022xpn} were computed
using the simple graviton gauge propagator
\cite{Tsamis:1992xa,Woodard:2004ut}. 
Nevertheless the secular logarithms persist even when 
computations are repeated using the 
graviton propagator in an exact de Sitter invariant gauge 
\cite{Miao:2011fc,Mora:2012zi}. When the vacuum 
polarization was recomputed in this covariant covariant 
gauge~\cite{Glavan:2015ura} and used to infer the 
corrections to the dynamical photon~\cite{Glavan:2016bvp}
the same secular correction was found, but with a different
coefficient. This indicates that the 
overall coefficients of large logarithms are
gauge dependent, and underscores the
need to isolate the physical part of logarithmically
enhanced quantum-gravitational corrections.

A formalism for isolating the physical long-distance
quantum-gravitational corrections to exchange potentials in
flat space is well developed. In momentum space, this
amounts to identifying the nonanalytic contributions to the
loop-corrected S-matrix
\cite{Donoghue:1994dn,Donoghue:1993eb,
Bjerrum-Bohr:2002aqa,Bjerrum-Bohr:2002gqz}. A position-space
generalization that avoids taking asymptotic limits of the
S-matrix was developed in
\cite{Miao:2017feh,Katuwal:2021thy} with the aim of 
extending the formalism
to de Sitter space, where asymptotic time limits typically
do not exist for massless fields. To define a physically meaningful massless scalar
exchange potential, this construction requires an effective
gauge-independent self-mass obtained by combining all
one-graviton-loop diagrams contributing to scattering
between a source and an observer, modeled by heavy scalar
fields~\cite{Glavan:2024elz}.

Although corrections to the external mode functions of the massive scalar field do not contribute to the long range force in flat space, recent calculations using a one-parameter family of gauges \cite{Glavan:2019msf,Glavan:2025azq} 
have shown that these corrections do matter in de Sitter \cite{Glavan:2026pug}.
To this end, we compute the one-graviton-loop correction to the
self-mass of a massive, minimally coupled scalar field in
de Sitter space. The calculation is dimensionally regulated,
and all divergences are absorbed into local BPHZ
counterterms, yielding a fully renormalized self-mass
 \cite{Bogoliubov:1957gp,Hepp:1966eg,
Zimmermann:1968mu,Zimmermann:1969jj}. Our techniques are
most closely related to previous self-mass computations, in the simple 
gauge~\cite{Tsamis:1992xa,Woodard:2004ut}, 
for a massless, minimally coupled scalar
\cite{Kahya:2007bc,Glavan:2021adm} and a massless, 
conformally coupled
scalar \cite{Glavan:2020gal}. The resulting self-mass 
we obtain is valid for arbitrary scalar mass and can therefore 
be used to study corrections, in the simple gauge~\cite{Tsamis:1992xa,Woodard:2004ut}, 
to the dynamics of both light and 
heavy fields, while resolving the
infrared and ultraviolet origin of secular effects.

The paper is organized as follows. In
Sec.~\ref{sec: Feynman rules}, we present the Feynman rules
and diagrams contributing to the scalar self-mass.
Section~\ref{sec: Self-mass in Minkowski space} summarizes
the corresponding Minkowski-space computation, which
introduces the computational strategy and provides a
consistency check for the de Sitter calculation. In
Sec.~\ref{sec: Propagators in de Sitter space}, we collect
the propagators used in
Sec.~\ref{sec: Evaluating one-loop diagrams in de Sitter space},
where the main computation is presented. In
Sec.~\ref{sec: Discussion}, we summarize the results and
discuss their implications. Additional identities and
technical details are collected in three appendices.

\section{Feynman rules}
\label{sec: Feynman rules}

We consider a massive scalar field~$\Psi$ minimally coupled
to gravity. The $D$-dimensional covariant action for this
system is
\begin{equation}
S[\Psi,g_{\mu\nu}]
    = \int\! d^{D\!}x \, \sqrt{-g} \,
    \bigg[ 
        \frac{R \!-\! (D\!-\!2)\Lambda}{\kappa^2}
        - \frac{1}{2} g^{\mu\nu}
            \partial_\mu \Psi \partial_\nu \Psi
        - \frac{m^2}{2} \Psi^2 \bigg]
        \, ,
\label{tree-level matter action}
\end{equation}
where~$\kappa^2\!=\!16\pi G_{\rm \scr N}$ is the rescaled 
Newton constant, and~$\Lambda\!=\!(D\!-\!1) H^2$ is the 
cosmological constant.
We work in the spatially flat cosmological patch of de
Sitter space, whose line element is
\begin{equation}
ds^2 = a^2(\eta) \big( - d\eta^2 + d\vec{x}^{\,2} \big) \, ,
\end{equation}
where the conformal time takes values
$\eta\!\in\!(-\infty,0)$ and the scale factor is
$a(\eta)\!=\!-1/(H\eta)$, with $H$ denoting the constant
physical Hubble rate.

The cubic and quartic interaction vertices required for our
calculation are obtained by expanding the action
in~(\ref{tree-level matter action}) after splitting the
metric into its de Sitter background and fluctuating parts,
\begin{equation}
g_{\mu\nu} = a^2 \big( \eta_{\mu\nu} + \kappa h_{\mu\nu} ) \, .
\end{equation}
The fluctuations defined in this way represent a
conformally rescaled graviton field, whose indices are
henceforth raised and lowered with the Minkowski metric
$\eta_{\mu\nu}$.
We represent the massive scalar by a dashed
line and the graviton by a wavy line. The corresponding
cubic and quartic interaction vertices are
\begin{align}
\vcenter{\hbox{\includegraphics{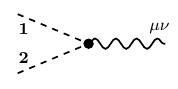}}}
={}&
    - i \kappa a^{D-2} 
        \bigg[ - \partial{}_1^{(\mu} \partial{}_2^{\nu)}
		+
        \frac{1}{2} \eta^{\mu\nu}
		\big( \partial{}_1 \!\cdot\! \partial{}_2 
            + a^2 m^2 \big)
		\bigg]
    \, ,
\label{3-vertex}
\\
\vcenter{\hbox{\includegraphics{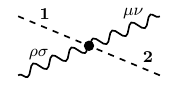}}}
    ={}&
	- i \kappa^2 a^{D-2}
	\biggl[
	\frac{1}{2} \partial{}_1^{(\mu} \eta^{\nu)(\rho} \partial{}_2^{\sigma)}
	+
	\frac{1}{2} \partial{}_1^{(\rho} \eta^{\sigma)(\mu} \partial{}_2^{\nu)}
	-
	\frac{1}{4} \partial{}_1^{(\mu} \partial{}_2^{\nu)} \eta^{\rho\sigma}
	-
	\frac{1}{4} \partial{}_1^{(\rho} \partial{}_2^{\sigma)} \eta^{\mu\nu}
\nonumber	\\
&
	\hspace{2.5cm}
	- \frac{1}{8} \bigl( 2 \eta^{\mu(\rho} \eta^{\sigma)\nu}  \!
	\!-\!
	\eta^{\mu\nu} \eta^{\rho\sigma} \bigr)
	\bigl( \partial{}_1 \!\cdot\! \partial{}_2 + a^2 m^2 \bigr)
	\biggr]
    \, .
\label{4-vertex}
\end{align}
Here $\partial_1$ and $\partial_2$ act on the two scalar
lines attached to the corresponding vertex, as indicated 
in the vertices.

\begin{figure}[h!]
\centering
\vskip+3mm
\hspace{2.cm}
\includegraphics[height=1.8cm]{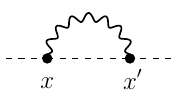}
\hfill
\includegraphics[height=1.8cm]{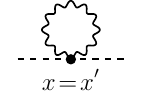}
\hfill
\includegraphics[height=1.8cm]{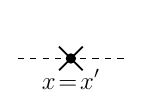}
\hspace{2.cm}
\vskip-3mm
\caption{One-loop diagrams representing contributions to the self-mass of the massive 
scalar field.
{\it Left:} the 3-vertex diagram. {\it Middle:} the 4-vertex diagram. {\it Right:} the counterterm diagram.}
\label{MassiveSelfMassDiagrams}
\end{figure}

The one-loop contributions to the scalar self-mass are shown
in Fig.~\ref{MassiveSelfMassDiagrams}.
For a spatially flat cosmological background, the
contributions from the 3-vertex and 4-vertex diagrams are
\begin{align}
-i\mathcal{M}^2_{\rm 3pt}(x;x')
	={}&
    - \kappa^2 (aa')^{D-2}
	 \bigg[ \overline{\partial}{}^{(\mu} \partial{}^{\nu)}
		- \frac{1}{2} \eta^{\mu\nu}
		\bigl( \overline{\partial}\!\cdot\! \partial 
            + m^2a^2 \bigr) 
		\bigg]
\nonumber \\
&   \hspace{1cm}
	\times\!
    \bigg[ \overline{\partial}{}'^{(\rho} \partial{}'^{\sigma)}
		- \frac{1}{2} \eta^{\rho\sigma}
		\bigl( {\overline{\partial}}{}' \!\cdot\! \partial{}' 
            + m^2 a'{}^2 \bigr) 
		\bigg]
	\begin{pmatrix}
	i \big[ {}_{\mu\nu} \Delta_{\rho\sigma} \big](x;x')
	\\
	i \Delta_m(x;x')
	\end{pmatrix}
	\, ,
\label{defM3pt}
\\
-i\mathcal{M}^2_{\rm 4pt}(x;x')
	={}&
	-
	i\kappa^2 \delta^D(x\!-\!x') a^{D-2}
	\bigg[
	\frac{1}{2} \overline{\partial}{}^{(\mu} \eta^{\nu)(\rho} \overline{\partial}{}'^{\sigma)}
	+
	\frac{1}{2}
	\overline{\partial}{}'^{(\mu} \eta^{\nu)(\rho} \overline{\partial}{}^{\sigma)}
	-
	\frac{1}{4} \eta^{\mu\nu} \overline{\partial}{}^{(\rho} \overline{\partial}{}'^{\sigma)} 
\nonumber \\
&   \hspace{-1cm}
	-
	\frac{1}{4} \overline{\partial}{}^{(\mu} \overline{\partial}{}'^{\nu)} \eta^{\rho\sigma}
	-
	\frac{1}{8} \bigl( 2 \eta^{\mu(\rho} \eta^{\sigma)\nu}
	\!-\!
	\eta^{\mu\nu} \eta^{\rho\sigma} \bigr)
		\bigl( \overline{\partial} \!\cdot\! \overline{\partial}{}' + m^2 a^2 \bigr)
	\bigg]
	i \big[ {}_{\mu\nu} \Delta_{\rho\sigma} \big](x;x)
	\, ,
\label{defM4pt}
\end{align}
where we introduce a simplified bi-local notation
for the scale factor, $a \!=\! a(\eta)$, $a'\!=\!a(\eta')$, 
and for derivatives, $\partial_\mu \!=\! \frac{\partial}{\partial x^\mu}$,
$\partial'_\mu \!=\! \frac{\partial}{\partial {x'}^\mu}$,
and where parenthesized indices denote weighted symmetrization.
We use the convention that barred
derivatives act on the external amputated scalar legs,
and unbarred derivatives act on the massive scalar propagator 
inside the loop. In
Eq.~(\ref{defM3pt}), the product of the graviton and scalar propagators is
displayed vertically in parentheses to mimic their relative
positions in the first diagram of
Fig.~\ref{MassiveSelfMassDiagrams}.

The ultraviolet divergences generated by the loop diagrams
are absorbed by local counterterms, represented by the last
diagram in Fig.~\ref{MassiveSelfMassDiagrams}. The
counterterm action is
\begin{equation}
S_{\rm ctm} = 
	\int\! d^{D\!}x \, \sqrt{-g} \,
	\frac{\kappa^2}{2}
	\Big[
	\big( \alpha_1 m^2
		+ \beta_1 H^2 \big)
		(\partial \Psi )^2
	+
	\big( \alpha_2 m^2 + \beta_2  H^2 \big) m^2 \Psi^2
	+
	\beta_3 H^2 (\nabla \Psi )^2
	\Big]
	\, ,
\label{QGcounter}
\end{equation}
where~$(\partial\Psi)^2\!=\! g^{\mu\nu} \partial_\mu \Psi \partial_\nu \Psi$
and~$(\nabla\Psi)^2\!=\! g^{ij} \partial_i \Psi \partial_j \Psi$.
The corresponding counterterm contribution to the self-mass is
\begin{align}
- i \mathcal{M}^2_{\rm ctm}(x;x')
	={}&
	\kappa^2 \Big[
	\big( \alpha_1 m^2 + \beta_1 H^2 \big)
		(aa')^{\frac{D-2}{2}}
		\overline{\partial} \!\cdot\! \overline{\partial}{}'
\nonumber \\
&	\hspace{0.8cm}
	+
	\big( \alpha_2 m^2 + \beta_2 H^2 \big)
		(aa')^\frac{D}{2} m^2
	+
	\beta_3H^2 (aa')^{\frac{D-2}{2}} \overline{\nabla} \!\cdot\! \overline{\nabla}{}'
	\Big]
	i \delta^D(x\!-\!x')
	\, .
\label{SelfMassCtm}
\end{align}
Note that the~$\beta_n$ counterterms, that contain explicit 
factors of~$H^2$, fundamentally descend from non-minimally 
coupled counterterms, where the background dependence is expressed in terms of the Ricci 
scalar~$H^2 \!\to\! R/[D(D\!-\!1)]$. The counterterm
proportional to~$\beta_3$ in not covariant, but it is 
necessary for the de Sitter space computation in order
to absorb the divergences generated in 
the noncovariant graviton gauge, introduced later in 
Sec.~\ref{sec: Propagators in de Sitter space}.

\section{Self-mass in Minkowski space}
\label{sec: Self-mass in Minkowski space}

The one-graviton-loop self-energy of massive scalar matter in 
Minkowski space was previously studied using causal 
perturbation theory~\cite{Grillo:1999ew}, while the one-loop 
ultraviolet divergences of gravity coupled to scalar matter 
had already been determined in Ref.~\cite{tHooft:1974toh}. 
Here we repeat the
calculation in position space using dimensional
regularization, both to introduce the techniques needed in
de Sitter space and to provide a flat-space check of the
subsequent computation. The Minkowski-space limit of the 
self-energy diagrams~(\ref{defM3pt}) and~(\ref{defM4pt}) 
is obtained by simply setting all the scale factors to 
one,~$a\!=\!a'\!=\!1$, while the counterterm 
contribution~(\ref{SelfMassCtm}) in addition requires
setting explicitly~$H\!\to\!0$.

\paragraph{Massive scalar propagator.}
The position-space Feynman propagator of a massive scalar 
field satisfies the equation of motion
\begin{equation}
\big( \partial^2 \!-\! m^2 \big) i \Delta_m(x;x')
    =
    i \delta^D(x\!-\!x')
    \, .
\label{FlatMassiveEOM}
\end{equation}
Its Lorentz-invariant solution is
\begin{equation}
i\Delta_m(x;x') =
    \frac{m^{D-2}}{(2\pi)^{\frac{D}{2}} }
    \frac{K_{\frac{D-2}{2}} \big( m\sqrt{\Delta x^2} \, \big)}
        {\big( m\sqrt{\Delta x^2} \, \big)^{\!\frac{D-2}{2}} }
    \, ,
\end{equation}
where~$K_n$ is the modified Bessel function of order~$n$,
and~$\Delta x^2$ is the Lorentz-invariant distance,
\begin{equation}
\Delta x^2 = \big\|\vec{x} \!-\! \vec{x}^{\,\prime} \big\| ^2
    -
    \big( |t \!-\! t'| \!-\! i\varepsilon \big)^2 ,
\label{Deltax2Def}
\end{equation}
with the~$i\varepsilon$-prescription corresponding to the
Feynman propagator.
The power-series representation of this propagator close
to coincidence is
\begin{equation}
i \Delta_m =
    \frac{m^{D-2} \,\Gamma \big( \frac{D-2}{2} \big) }
        { (4\pi)^{\frac{D}{2}} }
    \bigg[
        \Big( \frac{m^2\Delta x^2}{4} \Big)^{ \! \frac{2-D}{2}}
    +
    f_m^0(\Delta x^2)
    \bigg]
    \, ,
\label{MassivePropExpan}
\end{equation}
where we define
\begin{equation}
f_m^N(\Delta x^2)
    \equiv 
    \sum_{n=N}^{\infty}
    \Big( \frac{m^2 \Delta x^2}{4} \Big)^{\!n}
    \bigg[
    \frac{\Gamma \big( \frac{4-D}{2} \big) }
        { \Gamma\big(\frac{6-D}{2} \!+\! n \big) \,(n\!+\!1)! }
        \Big( \frac{m^2\Delta x^2}{4} \Big)^{\! \frac{4-D}{2}}
    -
    \frac{\Gamma \big( \frac{4-D}{2} \big)}
        { \Gamma \big(\frac{D}{2}\!+\!n) \,n! } \
    \bigg]
    \, .
\label{FlatDefF}
\end{equation}
This decomposition separates the leading short-distance
singularity, which must be retained in $D$ dimensions in 
order to regulate the self-mass divergences, from
the less singular remainder. In particular, we will use the
$D\!\to\!4$ limit
\begin{equation}
f_m^0(\Delta x^2)
    \xrightarrow{D \to 4}
    \frac{4 K_1 \big( m\sqrt{ \Delta x^2 } \, \big) }
        {m\sqrt{ \Delta x^2 }}
    - \frac{4}{m^2 \Delta x^2}
    \, .
\label{FlatRestSeries}
\end{equation}

\paragraph{Graviton propagator.}
The $\alpha=1$, $\beta=0$ specialization of the two-parameter 
graviton propagator, derived by Capper~\cite{Capper:1979ej},
in position space reads
\begin{equation}
i \big[ {}_{\mu\nu} \Delta {}_{\rho\sigma} \big] (x;x')
    =
    \Bigl[
        2\eta_{\rho(\mu}\eta_{\nu)\sigma}
		- 
        \frac{ 2 \eta_{\mu\nu}\eta_{\rho\sigma} }{ D \!-\! 2 }
        \Bigr]
        i \Delta (x;x') \, ,
\label{FlatGravProp}
\end{equation}
where the massless scalar propagator is
\begin{equation}
i \Delta (x;x')
    =
    \frac{\Gamma\big( \frac{D-2}{2} \big)}
        {4\pi^{ \frac{D}{2} } \Delta x^{D-2}}
        \, ,
\end{equation}
which coincides with the first term in the massive-propagator 
expansion~(\ref{MassivePropExpan}). This propagator is the 
flat-space limit of the de Sitter graviton propagator 
we introduce in Sec.~\ref{sec: Propagators in de Sitter space} 
and use in Sec.~\ref{sec: Evaluating one-loop diagrams in de Sitter space}.

\paragraph{Scalar self-mass.}
The dimensionally-regulated coincidence limit of the graviton 
propagator vanishes,
\begin{equation}
i \Delta (x;x) = 0
\qquad \Longrightarrow \qquad
i \big[ {}_{\mu\nu} \Delta {}_{\rho\sigma} \big] (x;x) = 0
\, ,
\label{FlatCoincidence}
\end{equation}
because there are no scales in the massless scalar propagator.
This immediately implies that the 4-vertex diagram 
in~(\ref{defM4pt}) does not contribute to the self-mass,
\begin{equation}
- i \mathcal{M}_{\rm 4pt}^2(x;x') = 0 \, .
\end{equation}

To evaluate the 3-vertex diagram
in~(\ref{defM3pt}), we first substitute the graviton
propagator~(\ref{FlatGravProp}), and contract its tensor
structure,
\begin{align}
\MoveEqLeft[18]
-i\mathcal{M}^2_{\rm 3pt}(x;x')
	=
    - \kappa^2
    i \Delta(x;x')
	\bigg[ 
    ( \overline{\partial} \!\cdot\! \overline{\partial}{}' )
    ( \partial{} \!\cdot\! \partial{}' )
    +
    ( \overline{\partial} \!\cdot\! \partial' )
    ( \overline{\partial}{}' \!\cdot\! \partial )
    -
    ( \overline{\partial} \!\cdot\! \partial )
    ( \overline{\partial}{}' \!\cdot\! \partial' )
\nonumber \\
&
		-
        m^2
		\big( \overline{\partial}\!\cdot\! \partial 
            \!+\!
            \overline{\partial}{}' \!\cdot\! \partial{}' \big)
        - \frac{Dm^4}{D\!-\!2}
		\bigg]
        i \Delta_m(x;x')
	\, ,
\end{align}
Using the equation of motion~(\ref{FlatMassiveEOM}), the
relative-coordinate dependence of the propagators, and the
coincidence limit~(\ref{FlatCoincidence}), we can rewrite
this expression as
\begin{equation}
-i\mathcal{M}^2_{\rm 3pt}(x;x')
	=
    \kappa^2 m^2
    i \Delta(x;x')
	\Bigl[ 
        \overline{\partial} \!\cdot\! \overline{\partial}{}'
		+
		\overline{\partial}\!\cdot\! \partial 
            +
            \overline{\partial}{}' \!\cdot\! \partial{}'
		+
        \frac{Dm^2}{D\!-\!2}
		\Bigr]
        i \Delta_m(x;x')
	\, ,
\end{equation}

The massless scalar propagator contains a single power-law
term. Power counting about coincidence shows that only the
leading term in the massive-propagator
expansion~(\ref{MassivePropExpan}) can generate an
ultraviolet divergence. The remainder may therefore be
evaluated directly in $D\!=\!4$:
\begin{align}
-i\mathcal{M}^2_{\rm 3pt}(x;x')
	={}&
    \frac{\kappa^2 m^2 \,\Gamma^2\big( \frac{D-2}{2} \big)}
        {16\pi^D \Delta x^{D-2}}
	\Bigl[ 
        \overline{\partial} \!\cdot\! \overline{\partial}{}'
		+
		\overline{\partial}\!\cdot\! \partial 
            +
            \overline{\partial}{}' \!\cdot\! \partial{}'
		+
        \frac{Dm^2}{D\!-\!2}
		\Bigr]
        \frac{1}{\Delta x^{D-2} }
\nonumber \\
&
    +
    \frac{\kappa^2 m^4}{64\pi^4 \Delta x^2}
	\Bigl[ 
        \overline{\partial} \!\cdot\! \overline{\partial}{}'
		+
		\overline{\partial}\!\cdot\! \partial 
            +
            \overline{\partial}{}' \!\cdot\! \partial{}'
		+
       2m^2
		\Bigr]
    f_m^0
    \, .
\end{align}
We next extract the external derivatives. This is immediate
for the first line and can be accomplished for the second
line by introducing a primitive function:
\begin{align}
-i\mathcal{M}^2_{\rm 3pt}(x;x')
	={}&
    \frac{\kappa^2 m^2 \,\Gamma^2\big( \frac{D-2}{2} \big)}
        {16\pi^D}
	\bigg[ 
        \overline{\partial} \!\cdot\! \overline{\partial}{}'
		+
		\tfrac{1}{2}\bigl(\overline{\partial}\!\cdot\! \partial 
            +
            \overline{\partial}{}' \!\cdot\! \partial{}'
            \bigr)
		+
        \frac{Dm^2}{D\!-\!2}
		\bigg]
        \frac{1}{\Delta x^{2D-4} }
\nonumber \\
&
    +
    \frac{\kappa^2 m^4}{64\pi^4}
	\biggl[ 
        \big(
        \overline{\partial} \!\cdot\! \overline{\partial}{}'
		\!+\!
        2m^2
        \big)
        \frac{f_m^0}{\Delta x^2}
		+
		\big( \overline{\partial}\!\cdot\! \partial 
            +
            \overline{\partial}{}' \!\cdot\! \partial{}'
            \big)
            \Big(  \frac{f_m^0}{\Delta x^2}
        +
        \mathcal{I} \Big[ \frac{f_m^0}{\Delta x^4} \Big] \Big)
		\biggr]
    \, ,
\end{align}
where the primitive function with respect to $\Delta x^2$ is 
denoted by
\begin{equation}
\mathcal{I}\big[ h(\Delta x^2) \big] 
    \equiv \int^{\Delta x^2} \!\!\! \! ds \, h(s) \, ,
\end{equation}
with the constant of integration understood to be set to
zero in the power-series representation.

We then use the identity~(\ref{1stIDextraction})
which isolates the ultraviolet divergence. After integrating
by parts to reflect the derivatives onto the appropriate
structures, we obtain the regulated expression with a
localized divergence:
\begin{align}
-i\mathcal{M}^2_{\rm 3pt}(x;x')
	\ \overset{D\to4}{\longsim} \  {}&
    \kappa^2 m^2
	\Bigl[ 
        4\overline{\partial} \!\cdot\! \overline{\partial}{}'
		\!-\!
        (D\!-\!8) m^2
		\Bigr]
    \bigg[
    \frac{\Gamma\big( \frac{D-2}{2} \big) \, \mu^{D-4} \, i \delta^D(x\!-\!x') }
		{ 16\pi^{\frac{D}{2}}(D\!-\!3)(D\!-\!4) }
	+
	\frac{ \partial \!\cdot\! \partial' }{128\pi^4} 
		\bigg( \frac{ \ln(\mu^2\Delta x^2) }{ \Delta x^2 } \bigg)
    \bigg]
\nonumber \\
&
    +
    \frac{\kappa^2 m^4}{64\pi^4}
	\bigg[ 
        \big(
        3 \overline{\partial} \!\cdot\! \overline{\partial}{}'
		\!+\!
        2m^2
        \big)
        \frac{f_m^0}{\Delta x^2}
		+
		2 \overline{\partial} \!\cdot\! \overline{\partial}{}'
            \mathcal{I} \Big[ \frac{f_m^0}{\Delta x^4} \Big]
		\, \bigg]
    \, .
\end{align}
The divergent local terms are absorbed by the flat-space
counterterms in~(\ref{SelfMassCtm}),
\begin{equation}
\alpha_1 =
    -
    \frac{4 \widetilde{\mu}^{D-4} }{D\!-\!4}
    +
    \frac{\alpha_1^{\rm fin}}{16\pi^2}
    \, ,
\qquad\quad
\alpha_2 =
    \frac{ (D\!-\!8) \widetilde{\mu}^{D-4}}{D\!-\!4}
    +
    \frac{\alpha_2^{\rm fin}}{16\pi^2}
    \, ,
\end{equation}
where we have introduced the rescaled
renormalization scale
\begin{equation}
\widetilde{\mu}^{D-4} =
    \frac{\Gamma\big( \frac{D-2}{2} \big) \, \mu^{D-4} }
		{ 16\pi^{\frac{D}{2}}(D\!-\!3) }
    \, ,
\label{MuTildeDef}
\end{equation}
and where~$\alpha_1^{\rm fin}$ 
and~$\alpha_2^{\rm fin}$ are suitably rescaled finite 
parts of the counterterms that are not specified by 
renormalization but should be fixed by further 
renormalization conditions.

After adding the counterterm contribution and taking
$D\!\to\!4$, we obtain the final result
\begin{align}
-i\mathcal{M}^2_{\rm ren}(x;x')
	={}&
    \frac{\kappa^2 m^2}{64\pi^4}
    \bigg[
	2 \big(
        \partial \!\cdot\! \partial{}'
		\!+\! m^2 \big) \partial \!\cdot\! \partial'
        \bigg(
		\frac{ \ln(\mu^2\Delta x^2) }{ \Delta x^2 }
        \bigg)
    +
        m^2\big(
        3 \partial \!\cdot\! \partial{}'
		\!+\!
        2m^2
        \big)
        \frac{f_m^0}{\Delta x^2}
\nonumber \\
&   \hspace{2cm}
		+
		2m^2 \partial \!\cdot\! \partial{}'
            \mathcal{I} \Big[ \frac{f_m^0}{\Delta x^4} \Big]
        +
        4 \pi^2 \big(
	\alpha_1^{\rm fin} \partial \!\cdot\! \partial'
	+
	\alpha_2^{\rm fin} m^2
	\big)
	i \delta^4(x\!-\!x')
		\biggr]
    \, ,
\label{FlatSelfMass1}
\end{align}
In this expression all derivatives have been reflected to
act within the self-mass kernel. Their
normalizations have been rescaled for convenience.
The complete self-mass in this gravtion gauge
vanishes in the massless limit, which is consistent with the 
flat space limit of the previously reported computations
\cite{Kahya:2007bc,Glavan:2021adm}.

The primitive function in Eq.~(\ref{FlatSelfMass1}) can be
eliminated in favor of simpler functions by using
identity~(\ref{FlatIdentity}), proven in
Appendix~\ref{app: Additional useful identities}. 
The renormalized
self-mass can consequently be written in an equivalent form
\begin{align}
\MoveEqLeft[10]
-i\mathcal{M}^2_{\rm ren}(x;x')
	=
    \frac{\kappa^2 m^2}{64\pi^4}
    \bigg\{
	\big(
        2\partial \!\cdot\! \partial{}'
		\!+\! m^2 \big) 
        \bigg[
        \partial \!\cdot\! \partial'
        \bigg(
		\frac{ \ln(\mu^2\Delta x^2) }{ \Delta x^2 }
        \bigg)
        +
        \frac{m^2 f_m^0}{\Delta x^2}
        \bigg]
\nonumber \\
&
    +
    4 \pi^2 
    \bigg(
	\alpha_1^{\rm fin} \partial \!\cdot\! \partial'
	+
	\Big[ \ln\!\Big(\frac{m^2}{4\mu^2}\Big) 
        + 1 + 2\gamma_{\rm E} 
        + \alpha_2^{\rm fin} \Big] m^2
	\bigg)
	i \delta^4(x\!-\!x')
        \biggr\}
    \, .
\end{align}
This result agrees with its momentum space counterpart 
reported in~\cite{Grillo:1999ew}.

\section{Propagators in de Sitter space}
\label{sec: Propagators in de Sitter space}

The massive scalar and graviton propagators in de Sitter
space have a more intricate structure than their
Minkowski-space counterparts. This is due to the expansion
of space and the presence of a cosmological horizon, to
which both fields are sensitive because of their
nonconformal coupling to gravity. In this section, we
summarize the properties of the two propagators used to
compute the self-mass diagrams in~(\ref{defM3pt})
and~(\ref{defM4pt}).

As bilocal objects, the propagators are most conveniently
expressed in terms of the de Sitter-invariant distance
function
\begin{equation}
y(x;x') = H^2aa' \Delta x^2
        \, ,
\label{yDef}
\end{equation}
where~$\Delta x^2$ is defined in~(\ref{Deltax2Def}).
This distance function incorporates the
$i\varepsilon$-prescription appropriate for Feynman
propagators and thus corresponds to the~$(++)$ polarity of
the Schwinger-Keldysh formalism
\cite{Schwinger:1960qe,Mahanthappa:1962ex,
Bakshi:1962dv,Bakshi:1963bn,Keldysh:1964ud,
Chou:1984es,Jordan:1986ug,Ford:2004wc}. Propagators with the
other Schwinger-Keldysh polarities are obtained
straightforwardly from this one.

\subsection{Massive scalar propagator}

The propagator of a massive, minimally coupled scalar field
in de Sitter space obeys the equation of motion
\begin{equation}
\big( \mathcal{D} - m^2a^4 \big)i\Delta_m(x;x')
        = i \delta^D(x\!-\!x')
    \, ,
\label{scalar eom}
\end{equation}
where $\mathcal{D}\!\equiv\! \partial^\mu a^2 \partial_\mu
\!=\! a^2 \partial^2 \!-\! 2 Ha^3\partial_0$.
The solution to
Eq.~(\ref{scalar eom}) is well known and was originally
constructed by Chernikov and Tagirov~\cite{Chernikov:1968zm}:
\begin{equation}
i\Delta_m(x; x') 
    =
    \frac{ H^{D-2}}{(4\pi)^{\frac{D}{2}}}
        \frac{\Gamma \big( \frac{D-1}{2} \!+\! \nu \big) 
            \, \Gamma\big( \frac{D-1}{2} \!-\! \nu \big)}
                {\Gamma\left(\frac{D}{2}\right)} \,
    {}_{2}F_{1} \bigg(
        \Big\{ \frac{D\!-\!1}{2} \!+\! \nu,
        \frac{D\!-\!1}{2} \!-\! \nu \Big\} ,
        \Big\{ \frac{D}{2} \Big\} ,
        1\!-\!\frac{y}{4}
        \bigg)
    \, .
\label{deSitter-propagator-2F1}
\end{equation}
Here $y(x;x')$, the argument of the hypergeometric function
${}_2F_1$, is defined in~(\ref{yDef}), and the index
$\nu$ is given by
\begin{equation}
 \nu^2 = \Big( \frac{D\!-\!1}{2} \Big)^{\!2} - \frac{m^2}{H^2}
	\, .
\label{index nu}
\end{equation}
Expanding Eq.~(\ref{deSitter-propagator-2F1}) about the
light cone, $y\sim 0$, gives
\begin{equation}
i\Delta_m(x; x')
	=
	\frac{ H^{D-2} \, \Gamma\big(\tfrac{D-2}{2}\big) }{(4\pi)^{\frac{D}{2}}}
	\bigg[
	\Big( \frac{y}{4} \Big)^{\!\frac{2-D}{2}}
	\! +
	\mathcal{F}_m^0(y)
	\bigg]
	\, ,
\label{eq:deSitter-propagator-f0}
\end{equation}
where, for $N \!\ge\! 0$, we define the power series
\begin{align}
\mathcal{F}_m^N(y) \equiv{}&
	\frac{ \Gamma\big( \frac{4-D}{2} \big) }
		{ \Gamma\big( \frac{1}{2} \!+\! \nu \big) \, \Gamma\big( \frac{1}{2} \!-\! \nu \big) }
	\sum_{n=N}^{\infty}
	\bigg[
	\frac{ \Gamma\big( \frac{3}{2} \!+\! \nu \!+\! n \big) \, 
		\Gamma\big( \frac{3}{2} \!-\! \nu \!+\! n \big) }
			{ \Gamma\big( \frac{6-D}{2} \!+\! n \big) \, (n\!+\!1)! }
				\Big( \frac{y}{4} \Big)^{\! n - \frac{D-4}{2}}
\nonumber \\
&	\hspace{5cm}
	-
	\frac{ \Gamma\big( \frac{D-1}{2} \!+\! \nu \!+\! n \big) \,
		\Gamma\big( \frac{D-1}{2} \!-\! \nu \!+\! n \big) }
			{ \Gamma\big( \frac{D}{2} \!+\! n \big) \, n! }
				\Big( \frac{y}{4} \Big)^{\! n }
	\bigg]
	\, .
\label{Fdef}
\end{align}

We will not need the complete massive scalar propagator
in $D$ dimensions. Instead, it is sufficient to retain the
three terms that must be kept away from $D\!=\!4$:
\begin{equation}
i\Delta_m
	\ \overset{D\to4}{\longsim} \
	\frac{ H^{D-2} \, \Gamma\big(\tfrac{D-2}{2}\big) }{(4\pi)^{ \frac{D}{2} } }
	\bigg[
	\left(\frac{y}{4}\right)^{\! \frac{2-D}{2}}
	+
	C_m^2 \Big( \frac{y}{4} \Big)^{ \! \frac{4-D}{2}}
	+
	C_m^3
	+
	\mathcal{F}_m^1(y)
	\bigg]
	,
\label{MassiveExpansion}
\end{equation}
where the two coefficients are
\begin{equation}
C_m^2
	=
	\frac{ 4\nu^2 \!-\! 1 }{ 2(D\!-\!4) }
	\, ,
\qquad\quad
C_m^3
	=
	-
	\frac{ \Gamma\big( \frac{4-D}{2} \big) }
	{ \Gamma\big( \frac{D}{2} \big) }
	\frac{ \Gamma\big( \frac{D-1}{2} \!+\! \nu \big) \Gamma\big( \frac{D-1}{2} \!-\! \nu \big) }
		{ \Gamma\big( \frac{1}{2} \!+\! \nu \big) \, \Gamma\big( \frac{1}{2} \!-\! \nu \big) }
	\, ,
\label{MassiveScalarCoefficients}
\end{equation}
and $\mathcal{F}_m^1$ is assumed evaluated at $D\!=\!4$.

Ultimately, we express our results in terms of
$\mathcal{F}_m^0$ evaluated at $D\!=\!4$:
\begin{align}
\mathcal{F}_m^0 ={}&
    \Gamma \Big( \frac{3}{2} \!+\! \nu \Big) 
        \, \Gamma\Big( \frac{3}{2} \!-\! \nu \Big) \,
    {}_{2}F_{1} \Big(
        \Big\{ \frac{3}{2} \!+\! \nu ,
        \frac{3}{2} \!-\! \nu \Big\} ,
        \Big\{ 2 \Big\} ,
        1\!-\!\frac{y}{4}
        \Big)
    -
    \frac{4}{y}
\nonumber \\
={}&
    \frac{1}{\Gamma\big( \frac{1}{2} \!+\! \nu \big) \,       
        \Gamma\big( \frac{1}{2} \!-\! \nu \big)}
    \sum_{n=0}^{\infty}
	\frac{ \Gamma\big( \frac{3}{2} \!+\! \nu \!+\! n \big) \, 
		\Gamma\big( \frac{3}{2} \!-\! \nu \!+\! n \big) }
			{ n! \, (n\!+\!1)!}
    \Big( \frac{y}{4} \Big)^{\! n }
\nonumber \\
&	\hspace{2.5cm}
    \times\!
    \bigg[
    \ln\!\Big( \frac{y}{4} \Big)
    + \psi\Big( \frac{3}{2} \!+\! \nu \!+\! n \Big)
    + \psi\Big( \frac{3}{2} \!-\! \nu \!+\! n \Big)
    - \psi(1\!+\!n)
    - \psi(2\!+\!n)
	\bigg]
    \, .
\label{FpowerSeries}
\end{align}
This is accomplished using the relation
\begin{equation}
C_m^2 \Big(\frac{y}{4} \Big)^{ \! \frac{4-D}{2}}
	\!\! +
	C_m^3
	+
	\mathcal{F}_m^1(y)
    \xrightarrow{D\to4}
    \frac{m^2 \!-\! 2H^2}{H^2} \big[ \ln(y) + \Psi_m \big]
    +
    \mathcal{F}_m^1
    =
	\mathcal{F}_m^0
	\, ,
\label{F1toF0}
\end{equation}
where we have introduced the shorthand
\begin{equation}
\Psi_m = 2 \gamma_{\scr \rm E} - 2\ln(2) - 1 
		+ \psi\Big( \frac{3}{2} \!+\! \nu \Big)  + \psi\Big( \frac{3}{2} \!-\! \nu \Big)
		\, ,
\end{equation}
with~$\psi$ denoting the digamma function.
We will also use two four-dimensional identities that follow
from the propagator equation of motion~(\ref{scalar eom})
and the power-series
representation~(\ref{eq:deSitter-propagator-f0}):
\begin{align}
\big( \mathcal{D} - m^2 a^4 \big) 
    \Big( \frac{4}{y} \!+\!  \mathcal{F}_m^0 \Big)
	={}&
	\frac{16\pi^2}{H^2} i \delta^4(x\!-\!x')
	\, ,
\label{eqF}
\\
\big( \mathcal{D} - m^2 a^4 \big) \mathcal{F}_m^0
	={}&
	\frac{4a^4(m^2 \!-\! 2H^2)}{y}
	\, ,
\label{eqF0}
\\
\big( \mathcal{D} - m^2 a^4 \big) \mathcal{F}_m^1
	={}&
	\frac{( m^2 \!-\! 2 H^2 ) a^4}{H^2}
        \Big( m^2 \big[ \ln(y) + \Psi_m \big] + 3H^2 \Big)
	\, .
\label{eqF1}
\end{align}
%

\subsection{Graviton propagator}
\label{subsec: Graviton propagator}

The graviton propagator in the simple non-invariant gauge
\cite{Tsamis:1992xa,Woodard:2004ut} can be decomposed into
three parts,
\begin{equation}
i \big[ {}_{\mu\nu} \Delta_{\rho\sigma} \big](x;x')
	=
	\sum_{N=I, I\!I, I\!I\!I}^{} \! t^N_{\mu\nu\rho\sigma} \!\times\! \mathcal{A}_N(x;x')
	\, .
\label{GravitonExpansion}
\end{equation}
Each part contains a constant tensor structure,
\begin{equation}
t_{\mu\nu\rho\sigma}^I
	=
	2\eta_{\rho(\mu}\eta_{\nu)\sigma}
	-
	\frac{ 2\eta_{\mu\nu}\eta_{\rho\sigma} }{ D \!-\! 2 }
	\, ,
\qquad
t_{\mu\nu\rho\sigma}^{I\!I}
	=
	2 \overline{\eta}_{\rho(\mu} \overline{\eta}_{\nu)\sigma}
		- \frac{2 \overline{\eta}_{\mu\nu} \overline{\eta}_{\rho\sigma} }{ D \!-\! 3 }
	\, ,
\qquad
t_{\mu\nu\rho\sigma}^{I\!I\!I}
	=
	- 4 \delta^0_{(\mu} \overline{\eta}_{\nu)(\rho} \delta^0_{\sigma)}
	\, ,
\end{equation}
where~$\overline{\eta}_{\mu\nu} \!=\! \eta_{\mu\nu} \!+\! \delta_\mu^0 \delta_\nu^0$. Each part also contains a scalar 
structure function expressed in terms of
scalar propagators,
\begin{equation}
\mathcal{A}_I = i\Delta_C \, ,
\qquad \qquad
\mathcal{A}_{I\!I} = i\Delta_A - i\Delta_C \, ,
\qquad\qquad
\mathcal{A}_{I\!I\!I} = i\Delta_B - i\Delta_C \, .
\label{StructureFunctions}
\end{equation}
The $A$-, $B$-, and $C$-type scalar propagators appearing in 
the structure functions satisfy the equation of 
motion~(\ref{scalar eom}) with respective masses given by,
\begin{equation}
M_A^2 = 0
	\, ,
\qquad\quad
M_B^2 =
	(D\!-\!2) H^2
	\, ,
\qquad\quad
M_C^2 = 2(D\!-\!3) H^2
	\, .
\end{equation}
The~$A$-type propagator is the propagator for the massless, 
minimally coupled scalar \cite{Onemli:2002hr}, which is not de 
Sitter-invariant. By contrast, the $B$- and $C$-type
propagators reduce to a conformally coupled scalar propagator
in $D\!=\!4$ and both are de Sitter invariant.

This graviton propagator is particularly simple for two
reasons: its tensor structures are constant, and the
power-series representations~(\ref{Fdef}) of all three
scalar propagators entering the structure functions
terminate at $D\!=\!4$. We will not require the complete
graviton propagator in arbitrary $D$ dimensions, but only
the first few terms that are
relevant to the ultraviolet divergences near $D\!=\!4$
in the self-mass. All three structure functions 
are contained in the form
\begin{equation}
\mathcal{A}_N
	\ \overset{D\to4}{\longsim} \
	\frac{ H^{D-2} \, \Gamma\big(\tfrac{D-2}{2}\big) }{(4\pi)^{ \frac{D}{2} } }
	\bigg[
	C_N^1 \Big( \frac{y}{4} \Big)^{\! \frac{2-D}{2}}
	+
	C_N^2 \Big( \frac{y}{4} \Big)^{ \! \frac{4-D}{2}}
	+
	C_N^3
	+
	C_N^4 \ln(aa')
	\bigg]
	\, ,
\label{GeneralFormStructure}
\end{equation}
where the four coefficients are listed in Table~\ref{CoeffTable}, 
and the divergent constant appearing there is
\begin{equation}
\Psi = - \psi\Big( \frac{2\!-\!D}{2}\Big) 
	+ \psi\Big(\frac{D\!-\!1}{2}\Big)
	+ \psi(D\!-\!1)
	- \gamma_{\rm \scr E}
    \ \overset{D\to4}{\longsim} \
    -
    \frac{2}{D\!-\!4}
    +
    \frac{5}{2}
    -
    2\ln(2)
    -
    2\gamma_{\rm \scr E}
    \, .
\end{equation}
%
%
\begin{table}[h!]
\renewcommand{\arraystretch}{1.5}
\centering
\begin{tabular}{ w{c}{1.cm} w{c}{3cm} w{c}{3cm} w{c}{3cm} w{c}{3cm} } 
\hline
	$\ N$
	&
	$C_N^1$
	&
	$C_N^2$
	&
	$C_N^3$
	&
	$C_N^4$
\\
\hline\hline
	$\ I$
	&
	$1$
	&
	$-\tfrac{6-D}{2}$
	&
	$\frac{\Gamma(D-3)}{\Gamma(\frac{D}{2}) \, \Gamma(\frac{D-2}{2}) }$
	&
	$0$
\\
	$\, I\!I$
	&
	$0$
	&
	$\frac{4(D-3)}{D-4}$
	&
	$\frac{\Gamma(D-3) [ (D-2) (D-3) \Psi - 1 ] }{\Gamma(\frac{D}{2}) \, \Gamma(\frac{D-2}{2}) } $
	&
	$2$
\\
	$I\!I\!I$
	&
	$0$
	&
	$2$
	&
	$- \frac{ 2 \, \Gamma(D-3)}{ \Gamma^2(\frac{D-2}{2}) } $
	&
	$0$
\\[0.8ex]
\hline
\end{tabular}
\caption{Coefficients from Eq.~(\ref{GeneralFormStructure})
for the three scalar structure functions defined 
in~(\ref{StructureFunctions}). All coefficients are 
finite in~$D\!=\!4$, except for the coefficients~$C_{I\!I}^2$ and~$C_{I\!I}^3$, which diverge as~$\sim\!1/(D\!-\!4)$. 
Only the
second structure function has a non-vanishing coefficient
in the last column, which signals de Sitter breaking
coming from the~$A$-type scalar propagator.}
\label{CoeffTable}
\end{table}

\noindent
We will also require the dimensionally regulated coincidence
limit of the full graviton propagator,
\begin{equation}
i \big[ {}_{\mu\nu} \Delta_{\rho\sigma} \big](x;x)
	=
	\frac{ H^{D-2} \, \Gamma\big(\tfrac{D-2}{2}\big) }{(4\pi)^{ \frac{D}{2} } }
	\times\!\!
	\sum_{N=I, I\!I, I\!I\!I}^{} \! t^N_{\mu\nu\rho\sigma} \!\times\! 
		\Big[ C_N^3 + 2C_N^4 \ln(a) \Big]
	\, .
\label{CoincidentGraviton}
\end{equation}
%

\section{Evaluating one-loop diagrams in de Sitter space}
\label{sec: Evaluating one-loop diagrams in de Sitter space}

This section evaluates the one-graviton-loop scalar self-mass
in de Sitter space. We treat the 4-vertex and 3-vertex
diagrams separately, including the renormalization of each
contribution. The 4-vertex diagram is local because it
involves a single graviton propagator evaluated at
coincidence. We then consider the 3-vertex diagram, whose
evaluation is more involved because it contains both the
graviton and massive scalar propagators away from
coincidence. Finally, we combine the renormalized
contributions and verify that the result reproduces the
Minkowski-space result of Sec.~3 in the flat-space limit.

\subsection{4-vertex diagram}

The middle diagram in Fig.~\ref{MassiveSelfMassDiagrams}
is evaluated in two steps. First, we substitute the
coincidence limit of the graviton
propagator~(\ref{CoincidentGraviton}) into the expression
for the 4-vertex diagram~(\ref{defM4pt}) and perform the
tensor contractions:
\begin{align}
-i\mathcal{M}^2_{\rm 4pt}(x;x')
	={}&
		\frac{ \kappa^2 H^{D-2} \, \Gamma\big(\tfrac{D-2}{2}\big) }{(4\pi)^{ \frac{D}{2} } }
	\bigg\{
	-
	\Big[
	D C_{I\!I}^3
	-
	(D\!-\!2) C_{I\!I\!I}^3
	\Big]
	(aa')^{\frac{D-2}{2}}
	\overline{\nabla}{} \!\cdot\! \overline{\nabla}{}'
\nonumber \\
&
	+
	\frac{1}{4}
	\Big[
	(D\!+\!1) (D\!-\!4) C_I^3
	+
	D (D\!-\!1) C_{I\!I}^3
	-
	2 (D\!-\!1) C_{I\!I\!I}^3
	\Big]
	(aa')^{\frac{D-2}{2}}\overline{\partial} \!\cdot\! \overline{\partial}{}'
\nonumber \\
&
	+
	\frac{1}{4}
	\Big[
	(D\!+\!1) D  C_I^3
	+
	D (D\!-\!1) C_{I\!I}^3
	+
	2 (D\!-\!1) C_{I\!I\!I}^3
	\Big]
	(aa')^{\frac{D}{2}} m^2
\nonumber \\
&
	+
	4 a^2 \ln(a)
	\Bigl[
	3\bigl( \overline{\partial} \!\cdot\! \overline{\partial}{}' \!+\! a^2 m^2 \bigr)
	-
	4\overline{\nabla}{} \!\cdot\! \overline{\nabla}{}'
	\Bigr]
	\bigg\}
		i\delta^D(x\!-\!x') 
	\, ,
\end{align}
Second, we absorb the local divergences into the
counterterms given in~(\ref{SelfMassCtm}). In the massless
limit, these divergences agree with those found
in~\cite{Glavan:2021adm}. No flat-space divergences arise
from this diagram, so that
$\alpha_1^{\rm 4pt} \!=\! \alpha_2^{\rm 4pt} \!=\! 0$.
A convenient choice of the curvature-dependent counterterms,
including their finite parts, is
\begin{align}
\beta_1^{\rm 4pt}
	={}&
	\frac{ \Gamma\big(\tfrac{D-2}{2}\big) \, \mu^{D-4} }{(4\pi)^{ \frac{D}{2} } }
	\times
	-
	\frac{1}{4}
	\Big[
	(D\!+\!1) (D\!-\!4) C_I^3
	+
	D (D\!-\!1) C_{I\!I}^3
	-
	2 (D\!-\!1) C_{I\!I\!I}^3
	\Big]
		\, ,
\\
\beta_2^{\rm 4pt}
	={}&
	\frac{ \Gamma\big(\tfrac{D-2}{2}\big) \, \mu^{D-4} }{(4\pi)^{ \frac{D}{2} } }
	\times
	-
	\frac{1}{4}
	\Big[
	(D\!+\!1) D  C_I^3
	+
	D (D\!-\!1) C_{I\!I}^3
	+
	2 (D\!-\!1) C_{I\!I\!I}^3
	\Big]
	\, ,
\\
\beta_3^{\rm 4pt}
	={}&
	\frac{ \Gamma\big(\tfrac{D-2}{2}\big) \, \mu^{D-4} }{(4\pi)^{ \frac{D}{2} } }
	\times
	\Big[
	D C_{I\!I}^3
	-
	(D\!-\!2) C_{I\!I\!I}^3
	\Big]
	\, .
\end{align}
Only the coefficient $C_{I\!I}^3$ contains a divergence.
For convenience, we also absorb the accompanying finite
local terms into the finite parts of the counterterms,
which yields the compact renormalized result
\begin{equation}
\big[ -i\mathcal{M}^2_{\rm 4pt} \big]^{\rm ren}(x;x')
	=
	\frac{ \kappa^2 H^2 }{ 8 \pi^2 }
	\Bigl[
	3\bigl( \overline{\partial} \!\cdot\! \overline{\partial}{}' \!+\! m^2aa' \bigr)
	-
	4 \overline{\nabla}{} \!\cdot\! \overline{\nabla}{}'
	\Bigr]
	aa'\ln\!\Big( \frac{\mu^2 aa'}{H^2} \Big)
	i\delta^4(x\!-\!x') 
	\, ,
\label{4ptRenormalized}
\end{equation}
where we have used $a'\!=\!a$ on the support of the delta
function to write the result in a manifestly symmetric
form. Notice that this contribution vanishes in the
flat space limit,~$H\!\to\!0$, consistent with the findings
of Sec.~\ref{sec: Self-mass in Minkowski space}.

\subsection{3-vertex diagram}

Computing the 3-vertex diagram, depicted on the left in
Fig.~\ref{MassiveSelfMassDiagrams}, is considerably more
involved. We therefore organize the calculation into a
sequence of modular steps. Following the decomposition of
the graviton propagator~(\ref{GravitonExpansion}) introduced
in Sec.~\ref{sec: Propagators in de Sitter space}, we split
the 3-vertex diagram into the three contributions labeled
$I$, $I\!I$, and $I\!I\!I$,
\begin{equation}
-i\mathcal{M}^2_{\rm 3pt}(x;x')
	=
	\sum_{N=I,I\!I,I\!I\!I}^{} 
	\!\!
	\big[ -i\mathcal{M}^2_{\rm 3pt} (x;x') \big]_N
	\, ,
\end{equation}
and evaluate each separately. 

We further organize each
contribution into terms according to the number of
derivatives acting on the massive scalar propagator:
\begin{align}
\big[ -i\mathcal{M}^2_{\rm 3pt} (x;x') \big]_N
	={}&
		- \kappa^2 (aa')^{D-2}
	\mathcal{A}_N(x;x')
	\Big[
	\Theta_N^{\mu\nu\rho\sigma}
		\overline{\partial}{}_\mu \partial_\nu 
	    	\overline{\partial}{}'_{\rho} \partial{}'_{\sigma} 
	-
	\tfrac{1}{2} (ma')^2
	\Theta_N^{\mu\nu}
		\overline{\partial}{}_\mu
	\partial_\nu
\nonumber \\
&	\hspace{2.2cm}
	-
	\tfrac{1}{2} (ma)^2
	\Theta_N^{\rho\sigma}
    		\overline{\partial}{}'_{\rho}
	\partial{}'_{\sigma} 
	+
	\tfrac{1}{4} (ma)^2(ma')^{2}
	\Theta_N
	\Big]
	i \Delta_m(x;x')
	\, .
\label{3ptN}
\end{align}
The term proportional to
$\Theta_N^{\mu\nu\rho\sigma}$ contains two derivatives
acting on the massive scalar propagator, the two terms
proportional to $\Theta_N^{\mu\nu}$ contain one derivative,
and the term proportional to $\Theta_N$ contains no
derivatives acting on it. 
These constant tensors are obtained by contracting the
tensor structures of the 3-vertices with those of the
graviton propagator decomposition:
\begin{subequations}
\begin{align}
&
\Theta_N^{\mu\nu\rho\sigma} =
	\Bigl[ 
    	\eta^{\alpha(\mu} \eta^{\nu)\beta} -
			\tfrac{1}{2} \eta^{\mu\nu} \eta^{\alpha\beta} \Big]
	\Bigl[ 
    	\eta^{\omega(\rho} \eta^{\sigma)\lambda} -
			\tfrac{1}{2} \eta^{\rho\sigma} \eta^{\omega\lambda} \Big]
	t^N_{\alpha\beta\omega\lambda}
	\, ,
\\
&
\Theta_N^{\mu\nu} =
	\Bigl[ 
    	\eta^{\alpha(\mu} \eta^{\nu)\beta} -
			\tfrac{1}{2} \eta^{\mu\nu} \eta^{\alpha\beta} \Big]
	\eta^{\rho\sigma}
	t^N_{\alpha\beta\rho\sigma}
	\, ,
\\
&
\Theta_N = \eta^{\mu\nu} \eta^{\rho\sigma} t^N_{\mu\nu\rho\sigma}
\, .
\end{align}
\label{Thetas}%
\end{subequations}
The tensor structures resulting from these contractions are
given in Table~\ref{TensorStructures}.

The scalar factors appearing in Eq.~(\ref{3ptN}) are
obtained by multiplying the graviton structure
functions~(\ref{GeneralFormStructure}) by the massive scalar
propagator~(\ref{MassiveExpansion}) and its derivatives:
\begin{align}
\mathcal{A}_N i \Delta_m
	\ \overset{D\to4}{\longsim} \ {}&
	\frac{ H^{2D-4} \, \Gamma^2\big(\tfrac{D-2}{2}\big) }{(4\pi)^{ D } }
	\bigg\{
	C_N^1 \Big(\frac{y}{4} \Big)^{\! 2-D}
	+
	\frac{4C_N^1}{y} \mathcal{F}_m^0(y)
\nonumber \\
&
	-
	\Big[
	\frac{D\!-\!4}{2} C_N^2 \ln\Big( \frac{y}{4} \Big)
	- \big(C_N^2 + C_N^3 \big)
	- C_N^4 \ln(aa')
	\Big]
	\Big( \frac{4}{y} \!+\! \mathcal{F}_m^0 \Big)
	\bigg\}
	\, ,
\label{ANDm}
\\
\mathcal{A}_N \partial_\mu i \Delta_m
	\ \overset{D\to4}{\longsim} \ {}&
	\frac{ H^{2D-4} \, \Gamma^2\big(\tfrac{D-2}{2}\big) }{(4\pi)^{ D } }
	\bigg\{
	C_N^1
	\Big(\frac{y}{4}\Big)^{\! \frac{2-D}{2}} \!\partial_\mu\Big(\frac{y}{4}\Big)^{\! \frac{2-D}{2}}
	+
	C_N^1 C_m^2
	\Big( \frac{y}{4} \Big)^{\! \frac{2-D}{2}}
	\!\partial_\mu \Big(\frac{y}{4}\Big)^{ \! \frac{4-D}{2}}
\nonumber \\
&
	+
	C_N^2 \Big[ \Big(\frac{y}{4}\Big)^{ \! \frac{4-D}{2}} \!\! - 1 \Big]
	\partial_\mu\Big(\frac{y}{4}\Big)^{\! \frac{2-D}{2}}
	-
	\frac{D\!-\!4}{2} C_N^2 \ln\Big( \frac{y}{4} \Big) \partial_\mu \mathcal{F}_m^0
\nonumber \\
&
	+
	\frac{4C_N^1}{y} \partial_\mu \mathcal{F}_m^1
	+
	\Big[ \big( C_N^2 + C_N^3 \big) + C_N^4 \ln(aa') \Big] \partial_\mu 
		\Big( \frac{4}{y} \!+\! \mathcal{F}_m^0 \Big)
	\bigg\}
	\, ,
\label{ANdDm}
\\
\mathcal{A}_N \partial_\mu \partial{}'_\nu i \Delta_m
	\ \overset{D\to4}{\longsim} \ {}&
	\frac{ H^{2D-4} \, \Gamma^2\big(\tfrac{D-2}{2}\big) }{(4\pi)^{ D } }
	\bigg\{
	C_N^1
	\Big(\frac{y}{4}\Big)^{\! \frac{2-D}{2}}
	\!\partial_\mu\partial{}'_\nu\Big(\frac{y}{4}\Big)^{\! \frac{2-D}{2}}
	+
	C_N^1 C_m^2 \Big( \frac{y}{4} \Big)^{\! \frac{2-D}{2}}
	\!\partial_\mu\partial{}'_\nu \Big(\frac{y}{4}\Big)^{ \! \frac{4-D}{2}}
\nonumber \\
&
	+
	C_N^2 \Big[ \Big(\frac{y}{4}\Big)^{ \! \frac{4-D}{2}} \!\! - 1 \Big]
	\partial_\mu \partial{}'_\nu\Big(\frac{y}{4}\Big)^{\! \frac{2-D}{2}}
	-
	\frac{D\!-\!4}{2} C_N^2 \ln\Big( \frac{y}{4} \Big) \partial_\mu\partial{}'_\nu \mathcal{F}_m^0
\nonumber \\
&
	+
	\frac{4C_N^1}{y} \partial_\mu\partial{}'_\nu 
        \mathcal{F}_m^1
	+
	\Big[ \big( C_N^2 + C_N^3 \big) + C_N^4 \ln(aa') \Big]
	\partial_\mu\partial{}'_\nu 
        \Big( \frac{4}{y} \!+\! \mathcal{F}_m^0 \Big)
	\bigg\}
	\, .
\label{ANddDm}
\end{align}
Note that we have kept only the non-integrable terms in arbitrary~$D$ dimensions, 
and that the rest are evaluated in~$D\!=\!4$ immediately.

\begin{table}[h!]
\vskip+3mm
\renewcommand{\arraystretch}{1.8}
\center
\begin{tabular}{ w{c}{1cm} w{c}{6cm} w{c}{4.5cm} w{c}{2.5cm} } 
\hline
	$N$
	&
	$\Theta_N^{\mu\nu\rho\sigma}$
	&
	$\Theta_N^{\mu\nu}$
	&
	$\Theta_N$
\\
\hline\hline
	$I$
	&
	$
    	2 \eta^{\rho(\mu} \eta^{\nu)\sigma} - \eta^{\mu\nu} \eta^{\rho\sigma}
	$
	&
	$2 \eta^{\mu\nu}$
	&
	$- \frac{4D}{D-2} $
\\
	$I\!I$
	&
	$
    2 \overline{\eta}^{\rho(\mu} \overline{\eta}^{\nu)\sigma}
	-
	\eta^{\mu\nu} \eta^{\rho\sigma}
	-
	\frac{2 }{ D-3 } 
    \delta^\mu_0 \delta^\nu_0 \delta^\rho_0 \delta^\sigma_0
	$
	&
	$ \frac{2(D-1)}{D-3} \eta^{\mu\nu} 
		\!-\! \frac{4  }{D-3} \overline{\eta}^{\mu\nu} $
	&
	$- \frac{4 (D-1) }{D-3}$
\\
	$I\!I\!I$
	&
	$
	-4 \delta{}_0^{(\mu} \overline{\eta}{}^{\nu)(\rho} \delta{}_0^{\sigma)}
	$
	&
	$0$
	&
	$0$
\\
\hline
\end{tabular}
\caption{Results for the tensor contractions 
in~(\ref{Thetas}).}
\label{TensorStructures}
\end{table}

In the remainder of this subsection, we first extract the
external derivatives from the nonintegrable short-distance
terms and express their ultraviolet divergences as local
distributions. We then contract the resulting coefficient
functions with the tensors in
Table~\ref{TensorStructures} and absorb the local
divergences into the available counterterms. The remaining
finite nonlocal terms are simplified by further extracting
external derivatives and applying integration-by-parts
identities.

\subsubsection{Localizing divergences}

We next extract derivatives from the nonintegrable
short-distance terms and express their ultraviolet
divergences as local distributions. This is accomplished
using the fundamental distributional identities collected
in Appendix~\ref{app: Localizing divergences}. 
From these, we derive the identities needed
for terms containing no derivatives,
\begin{equation}
\Big( \frac{H^2aa'}{4} \Big)^{\!D-2}
\Big(\frac{y}{4} \Big)^{\! 2-D}
	\ \overset{D\to4}{\longsim} \ 
	\frac{ 2\pi^{\frac{D}{2} } \mu^{D-4} i \delta^D(x\!-\!x') }
		{(D\!-\!3)(D\!-\!4) \, \Gamma\big( \frac{D-2}{2} \big)}
	+
	\frac{\partial \!\cdot\! \partial'}{4}
		\bigg( \frac{ \ln(\mu^2 \Delta x^2) }{ \Delta x^2 } \bigg)
		\, ,
\end{equation}
for terms containing one derivative,
\begin{align}
\Big( \frac{H^2aa'}{4} \Big)^{\!D-2}
\Big(\frac{y}{4}\Big)^{\! \frac{2-D}{2}} \!\partial_\mu\Big(\frac{y}{4}\Big)^{\! \frac{2-D}{2}}
	\ \overset{D\to4}{\longsim} \ {}&
	\big[ \partial_\mu \!-\! \delta_\mu^0 (D\!-\!2) aH \big]
	\frac{ \pi^{\frac{D}{2} } \mu^{D-4} i \delta^D(x\!-\!x') }
		{(D\!-\!3)(D\!-\!4) \, \Gamma\big( \frac{D-2}{2} \big)}
\nonumber \\
	& 
	+
	\frac{1}{8} 
    \big( \partial_\mu \!-\! 2\delta_\mu^0 aH \big) 
    \partial \!\cdot\! \partial'
		\bigg( \frac{ \ln(\mu^2 \Delta x^2) }{ \Delta x^2 } \bigg)
	\, ,
\\
\Big( \frac{H^2aa'}{4} \Big)^{\!D-2}
\Big( \frac{y}{4} \Big)^{\! \frac{2-D}{2}}
	\! \partial_\mu \Big(\frac{y}{4}\Big)^{ \! \frac{4-D}{2}}
	\ \overset{D\to4}{\longsim} \ {}&
	\frac{(D\!-\!4)}{8} H^4(aa')^2 \partial_\mu \frac{1}{y}
	\, ,
\\
\Big( \frac{H^2aa'}{4} \Big)^{\!D-2}
\Big[ \Big(\frac{y}{4}\Big)^{ \! \frac{4-D}{2}} \!\! - 1 \Big]
	\partial_\mu\Big(\frac{y}{4}\Big)^{\! \frac{2-D}{2}}
	\ \overset{D\to4}{\longsim} \ {}&
	-
	\frac{(D\!-\!4)}{8} H^4(aa')^2 \partial_\mu
	\bigg(
	\frac{ 1 + \ln \!\big( \frac{y}{4} \big) }{ y }
	\bigg)
	\, ,
\end{align}
and for terms containing two derivatives,
\begin{align}
&
\Big( \frac{H^2aa'}{4} \Big)^{\!D-2}
\Big(\frac{y}{4}\Big)^{\! \frac{2-D}{2}} \!\partial_\mu\partial{}'_\nu
	\Big(\frac{y}{4}\Big)^{\! \frac{2-D}{2}}
	\ \overset{D\to4}{\longsim} \ 
	\bigg[
	\frac{ D \partial_\mu \partial{}'_\nu - \eta_{\mu\nu} \partial \!\cdot\! \partial' }{D\!-\!1}
	-
	(D\!-\!2) \big( Ha \delta_\mu^0 \partial_\nu' + Ha' \delta_\nu^0 \partial_\mu \big)
\nonumber \\
&	\hspace{2.2cm}
	+
	(D\!-\!2)^2 H^2 aa' \delta_\mu^0 \delta_\nu^0
	\bigg]
	\bigg[
	\frac{ \pi^{\frac{D}{2} }  \mu^{D-4} \, i \delta^D(x\!-\!x') }
		{ 2(D\!-\!3)(D\!-\!4) \, \Gamma\big( \frac{D-2}{2} \big) }
	+
	\frac{ \partial \!\cdot\! \partial' }{16}
	\bigg( \frac{ \ln(\mu^2\Delta x^2) }{ \Delta x^2 } \bigg)
	\bigg]
	\, ,
\\
&
\Big( \frac{H^2aa'}{4} \Big)^{\!D-2} \Big( \frac{y}{4} \Big)^{\! \frac{2-D}{2}}
	\! \partial_\mu\partial{}'_\nu \Big(\frac{y}{4}\Big)^{ \! \frac{4-D}{2}}
	\ \overset{D\to4}{\longsim} \
	H^2aa' \eta_{\mu\nu}
	\frac{ \pi^{\frac{D}{2} }  \mu^{D-4} \, i \delta^D(x\!-\!x') }
		{ 4 (D\!-\!3) \, \Gamma\big( \frac{D-2}{2} \big) }	
\nonumber \\
&	\hspace{4cm}
	+
	\frac{(D\!-\!4)H^2aa'  }{32}
	\bigg[
	\eta_{\mu\nu} \partial \!\cdot\! \partial' \bigg(
		\frac{ \ln(\mu^2 \Delta x^2) }{ \Delta x^2 }
		\bigg)
	+
	2 \partial_\mu \partial{}'_\nu \frac{1}{\Delta x^2}
	\bigg]
	\, ,
\\
&
\Big( \frac{H^2aa'}{4} \Big)^{\!D-2}
\Big[ \Big(\frac{y}{4}\Big)^{ \! \frac{4-D}{2}} \!\! - 1 \Big]
	\partial_\mu\partial{}'_\nu \Big(\frac{y}{4}\Big)^{\! \frac{2-D}{2}}
	\ \overset{D\to4}{\longsim} \ \
	H^2aa' \eta_{\mu\nu} 
	\frac{\pi^{\frac{D}{2} }  \mu^{D-4} \, i \delta^D(x\!-\!x') }
		{ 4(D\!-\!3) \, \Gamma\big( \frac{D-2}{2} \big) }
\nonumber \\
&	\hspace{0.7cm}
	+
	\frac{(D\!-\!4)}{8} H^2aa'
	\bigg[
	\eta_{\mu\nu} 
	\frac{ \partial \!\cdot\! \partial'}{4} \bigg(
		\frac{ \ln(\mu^2 \Delta x^2) }{ \Delta x^2 }
		\bigg)
	+
	\frac{\partial_\mu \partial{}'_\nu}{2} \frac{ 1 }{ \Delta x^2 }
	-
	H^2aa' \partial_\mu \partial{}'_\nu
	\bigg(
	\frac{ 2 + \ln\!\big( \frac{y}{4} \big) }{ y }
	\bigg)
	\bigg]
	\, .
\end{align}
Applying these identities to
Eqs.~(\ref{ANDm})--(\ref{ANddDm}), and substituting the
coefficients of the massive scalar propagator from
Eq.~(\ref{MassiveScalarCoefficients}) and those of the
graviton propagator from Table~\ref{CoeffTable}, isolates
the nonintegrable terms and expresses their ultraviolet
divergences as local distributions. For the coefficient
functions containing no derivatives, we obtain
\begin{align}
(aa')^{D-2} \mathcal{A}_I i \Delta_m
	\ \overset{D\to4}{\longsim} \ {}&
	\frac{ \Gamma\big(\tfrac{D-2}{2}\big) \mu^{D-4} i \delta^D(x\!-\!x') }
		{ 8 \pi^{ \frac{D}{2} } (D\!-\!3)(D\!-\!4) }
	\!+
	\frac{\partial \!\cdot\! \partial'}{64 \pi^4}
		\bigg( \frac{ \ln(\mu^2 \Delta x^2) }{ \Delta x^2 } \bigg)
	\!+\!
	\frac{H^2aa'}{64 \pi^4\Delta x^2} \mathcal{F}_m^0
	\, ,
\\
(aa')^{D-2} \mathcal{A}_{I\!I} i \Delta_m
	\ \overset{D\to4}{\longsim} \ {}&
	\frac{H^4(aa')^2}{128\pi^4} 
	\big[ 2 \!-\! 2\gamma_{\rm \scr E} 
        \!-\! \ln( H^2 \Delta x^2 ) \big]
	\Big( \frac{4}{y} \!+\! \mathcal{F}_m^0 \Big)
	\, .
\end{align}
No $I\!I\!I$ coefficient functions with zero or one
derivative are required because the corresponding
tensor structures vanish, as shown in
Table~\ref{TensorStructures}.
For the coefficient functions containing one derivative, 
we find
\begin{align}
(aa')^{D-2} \mathcal{A}_I \partial_\mu i \Delta_m
	\ \overset{D\to4}{\longsim} \
	{}&
	\Big[ \partial_\mu - \delta_\mu^0 (D\!-\!2) aH \Big]
	\bigg[
	\frac{ \Gamma\big(\tfrac{D-2}{2}\big) \mu^{D-4} 
	 i \delta^D(x\!-\!x') }
		{ 16 \pi^{ \frac{D}{2} } (D\!-\!3)(D\!-\!4) }
	+
	\frac{ \partial \!\cdot\! \partial' }{ 128 \pi^4 }
		\bigg( \frac{ \ln(\mu^2 \Delta x^2) }{ \Delta x^2 } \bigg)
	\bigg]
\nonumber \\
&
	-
	\frac{ (m^2 \!-\! 2H^2) H^2 (aa')^2 }{ 64 \pi^4 }
	\partial_\mu \frac{1}{y}
	+
	\frac{H^4(aa')^2}{ 64 \pi^4 y}
	\partial_\mu \mathcal{F}_m^1
	\, ,
\\
(aa')^{D-2} \mathcal{A}_{I\!I} \partial_\mu i \Delta_m
	\ \overset{D\to4}{\longsim} \ {}&
	-
	\frac{H^4 (aa')^2}{32 \pi^4 }
	\partial_\mu
	\Big(
	\frac{ \ln ( y ) }{ y }
	\Big)
	+
	\frac{H^4 (aa')^2}{32 \pi^4 }
	\big[ 1 \!-\! 2\gamma_{\rm \scr E} 
        \!+\! \ln(aa') \big]
    \partial_\mu \frac{1}{y}
\nonumber \\
&
	+
	\frac{H^4 (aa')^2}{128 \pi^4 }
	\big[ 2 \!-\! 2\gamma_{\rm \scr E} 
        \!-\! \ln(H^2\Delta x^2) \big] 
    \partial_\mu \mathcal{F}_m^0
	\, .
\end{align}
Finally, the coefficient functions containing two
derivatives are
\begin{align}
\MoveEqLeft[3]
(aa')^{D-2} \mathcal{A}_I \partial_\mu \partial{}'_\nu i \Delta_m
	\, \overset{D\to4}{\longsim} \
	\bigg[
	\frac{ D \partial_\mu \partial{}'_\nu \!-\! \eta_{\mu\nu} \partial \!\cdot\! \partial' }{D\!-\!1}
	-
	(D\!-\!2) \big( Ha \delta_\mu^0 \partial_\nu' \!+\! Ha' \delta_\nu^0 \partial_\mu \big)
	\!+
	(D\!-\!2)^2 H^2 aa' \delta_\mu^0 \delta_\nu^0
\nonumber \\
&
	-
	\big[ m^2 \!-\! (D\!-\!2) H^2 \big] aa' \eta_{\mu\nu}
	\bigg]
	\bigg[
	\frac{ \Gamma\big(\tfrac{D-2}{2}\big) \, \mu^{D-4} \, i \delta^D(x\!-\!x') }
		{ 32 \pi^{ \frac{D}{2} } (D\!-\!3)(D\!-\!4) }
	+
	\frac{ \partial \!\cdot\! \partial' }{256 \pi^4}
	\bigg( \frac{ \ln(\mu^2\Delta x^2) }{ \Delta x^2 } \bigg)
	\bigg]
\nonumber \\
&
	-
	\frac{( m^2 \!-\! 2H^2 )aa'}{128\pi^4} \partial_\mu \partial{}'_\nu \frac{1}{\Delta x^2}
	+
	\frac{H^2 aa'}{ 64 \pi^4\Delta x^2} \partial_\mu\partial{}'_\nu \mathcal{F}_m^1
	\, ,
\\
\MoveEqLeft[3]
(aa')^{D-2} \mathcal{A}_{I\!I} \partial_\mu \partial{}'_\nu i \Delta_m
	\ \overset{D\to4}{\longsim} \
	H^2aa' \eta_{\mu\nu} 
	\bigg[
	\frac{ \Gamma\big( \frac{D-2}{2} \big) \, \mu^{D-4} \, i \delta^D(x\!-\!x') }
		{ 16 \pi^{ \frac{D}{2} }(D\!-\!4) }
	+
	\frac{ \partial \!\cdot\! \partial' }{ 128 \pi^4 } \bigg(
		\frac{ \ln(\mu^2 \Delta x^2) }{ \Delta x^2 }
		\bigg)
	\bigg]
\nonumber \\
&
	+
	\frac{ H^2aa' }{ 64 \pi^4 } \partial_\mu \partial{}'_\nu \frac{ 1 }{ \Delta x^2 }
	+
	\frac{H^4 (aa')^2}{128\pi^4}
	\big[ 2 \!-\! 2\gamma_{\rm \scr E} \!+\! \ln(aa') \big]
	\partial_\mu\partial{}'_\nu 
    \Big( \frac{4}{y} \!+\! \mathcal{F}_m^0 \Big)
\nonumber \\
&
	-
	\frac{ H^4(aa')^2 }{ 32 \pi^4 }
	\partial_\mu \partial{}'_\nu
	\Big(
	\frac{ 2+\ln( y) }{ y }
	\Big)
	-
	\frac{H^4 (aa')^2}{128\pi^4}
	\ln(y)
	\partial_\mu\partial{}'_\nu \mathcal{F}_m^0
	\, ,
\\
\MoveEqLeft[3]
(aa')^{D-2} \mathcal{A}_{I\!I\!I} \partial_\mu \partial{}'_\nu i \Delta_m
	\ \overset{D\to4}{\longsim} \
	H^2aa' \eta_{\mu\nu} \frac{i \delta^4(x\!-\!x') }{ 32 \pi^2 }
	\, .
\end{align}
These expressions separate the local ultraviolet terms from
the finite nonlocal remainders while retaining the external
derivative operators. We next contract them with the tensor
structures in Table~\ref{TensorStructures}.

\subsubsection{Contracting tensor structures}
\label{subsubsec: Contracting tensor structures}

After isolating the nonintegrable terms and localizing their
ultraviolet divergences, we contract the coefficient
functions and external derivatives with the tensor
structures in Table~\ref{TensorStructures}, and include the
remaining mass and scale-factor coefficients from
Eq.~(\ref{3ptN}). The derivative identities used throughout
this calculation are collected in
Appendix~\ref{subapp: Identities used with tensor contractions}.

We begin with the contributions associated with the
rightmost column of Table~\ref{TensorStructures}, which
contain no derivatives acting on the massive scalar
propagator:
\begin{align}
- \frac{\kappa^2}{4} \Theta_I m^4
	(aa')^{D} \mathcal{A}_I i \Delta_m
	\ \overset{D\to4}{\longsim} \ {}&
	\kappa^2 m^4 (aa')^2
	\bigg[
	\frac{ D \,\Gamma\big(\tfrac{D-2}{2}\big) \mu^{D-4} i \delta^D(x\!-\!x') }
		{ 8 \pi^{ \frac{D}{2} } (D\!-\!2) (D\!-\!3)(D\!-\!4) }
\nonumber \\
&	\hspace{2.5cm}
	+
	\frac{\partial \!\cdot\! \partial'}{32 \pi^4}
		\bigg( \frac{ \ln(\mu^2 \Delta x^2) }{ \Delta x^2 } \bigg)
	+
	\frac{H^4(aa')^2}{32 \pi^4 y} \mathcal{F}_m^0
	\bigg]
	\, ,
\label{Contracted0I}
\\
- \frac{\kappa^2}{4} \Theta_{I\!I} m^4
	(aa')^{D} \mathcal{A}_{I\!I} i \Delta_m
	\ \overset{D\to4}{\longsim} \ {}&
	\frac{3\kappa^2m^4H^4(aa')^4}{128\pi^4} 
	\big[
	2 \!-\! 2\gamma_{\rm \scr E} \!-\! \ln( H^2 \Delta x^2 )
	\big]
	\Big(
	\frac{4}{y}
	\!+\! 
	\mathcal{F}_m^0
	\Big)
	\, .
\label{Contracted0II}
\end{align}

The contributions associated with the middle column of
Table~\ref{TensorStructures} contain one derivative acting
on the massive scalar propagator. We use
Eqs.~(\ref{LocalId1}) and~(\ref{LocalId2}) to reorganize
the derivatives in the local terms, 
Eq.~(\ref{ReflSymmId1}) to reorder 
derivatives in the nonlocal terms,
and Eq.~(\ref{RedId1}) to reduce the derivative order of 
the nonlocal terms:
\begin{align}
\MoveEqLeft[5]
\frac{\kappa^2}{2}
\Theta_I^{\mu\nu} (aa')^{D-2} \mathcal{A}_I 
	\Big[ (ma')^2\overline{\partial}{}_\mu\partial{}_\nu
		+ (ma)^2 \overline{\partial}{}'_\mu\partial{}'_\nu \Big] i \Delta_m
\ \overset{D\to4}{\longsim} \ 
    \kappa^2 m^2
    aa' \Big[ 
        (D\!-\!4)(2m^2 \!-\! 7H^2)aa'
\nonumber \\
&
    +
    2\overline{\partial} 
        \!\cdot\! \overline{\partial}{}'
    \Big]
	\frac{ \Gamma\big(\tfrac{D-2}{2}\big) \mu^{D-4} 
	i \delta^D(x\!-\!x') }
		{ 16 \pi^{ \frac{D}{2} } (D\!-\!3)(D\!-\!4) }
    -
    \frac{\kappa^2 m^2}{(aa')^2}
    \big( a'^4 \overline{\mathcal{D}} + a^4 \overline{\mathcal{D}}{}'
		\big) 
	\frac{ \partial \!\cdot\! \partial' }{ 128 \pi^4 }
		\bigg( \frac{ \ln(\mu^2 \Delta x^2) }{ \Delta x^2 } 
        \bigg)
\nonumber \\
&
	+
	\frac{ \kappa^2 m^2 (m^2 \!-\! 2H^2) H^2 (aa')^3}
        { 16 \pi^4 \Delta x^2 }
	+
	\frac{\kappa^2 m^2 H^4(aa')^2}{ 64 \pi^4 y}
	\big( a'^2 \, \overline{\partial}{} \!\cdot\! \partial 
        + a^2 \, \overline{\partial}{}' \!\cdot\! \partial' \big)
    \mathcal{F}_m^1
    \, ,
\label{Contracted1I}
\\
\MoveEqLeft[5]
\frac{\kappa^2}{2}
\Theta_{I\!I}^{\mu\nu} (aa')^{D-2} \mathcal{A}_{I\!I} 
	\Big[ (ma')^2\overline{\partial}{}_\mu\partial{}_\nu
		+ (ma)^2 \overline{\partial}{}'_\mu\partial{}'_\nu \Big] i \Delta_m
\nonumber \\
	\ \overset{D\to4}{\longsim} \ {}&
	-
	\frac{\kappa^2 m^2 H^4 (aa')^2}{32 \pi^4 }
	\Big[
	a'{}^2 \big( 3 \overline{\partial}{} \!\cdot\! \partial{}
		\!-\! 2 \overline{\nabla}{} \!\cdot\! \nabla{} \big) 
	+
	a^2 \big( 3 \overline{\partial}{}' \!\cdot\! \partial{}'
		\!-\! 2 \overline{\nabla}{}' \!\cdot\! \nabla{}' \big)
	\Big]
	\frac{ 1 + \ln ( y ) }{ y }
\nonumber \\
&	\hspace{-0.8cm}
	-
	\frac{\kappa^2 m^2 H^4 (aa')^2}{128 \pi^4 }
	\ln(y)
	\Big[
	a'{}^2 \big( 3 \overline{\partial}{} \!\cdot\! \partial{}
		\!-\! 2 \overline{\nabla}{} \!\cdot\! \nabla{} \big) 
	+
	a^2 \big( 3 \overline{\partial}{}' \!\cdot\! \partial{}'
		\!-\! 2 \overline{\nabla}{}' \!\cdot\! \nabla{}' \big)
	\Big] 
	\mathcal{F}_m^0
\label{Contracted1II}
\\
&	\hspace{-1.5cm}
	+
	\frac{\kappa^2 m^2 H^4 (aa')^2}{128 \pi^4 }
	\big[ 2 \!-\! 2\gamma_{\rm \scr E} \!+\! \ln(aa')\big] 
	\Big[
	a'{}^2 \big( 3 \overline{\partial}{} \!\cdot\! \partial{}
		\!-\! 2 \overline{\nabla}{} \!\cdot\! \nabla{} \big) 
	+
	a^2 \big( 3 \overline{\partial}{}' \!\cdot\! \partial{}'
		\!-\! 2 \overline{\nabla}{}' \!\cdot\! \nabla{}' \big)
	\Big] 
	\Big( \frac{4}{y} \!+\! \mathcal{F}_m^0 \Big)
	\, .
\nonumber 
\end{align}

The remaining contributions are associated with the
leftmost column of Table~\ref{TensorStructures} and contain
two derivatives acting on the massive scalar propagator.
We use Eqs.~(\ref{LocalId3}) and~(\ref{LocalId4}) to
reorganize the derivatives in the local terms,
Eq.~(\ref{ReflSymmId2}) to reorder 
derivatives in the nonlocal terms, and Eqs.~(\ref{RedId2}) 
and~(\ref{RedId3}) to reduce the remaining derivative
structures:
\begin{align}
\MoveEqLeft[2]
- \kappa^2
\Theta_I^{\mu\nu\rho\sigma} (aa')^{D-2} \mathcal{A}_I 
	\overline{\partial}{}_\mu\partial{}_\nu \overline{\partial}{}'_\rho\partial{}'_\sigma i \Delta_m
\nonumber \\
&   \hspace{-0.3cm}
\ \overset{D\to4}{\longsim} \ 
	\kappa^2\Big[
	2 m^2 aa' \overline{\partial} 
        \!\cdot\! \overline{\partial}{}'
	-
	\frac{(D\!+\!2)}{2} H^2 aa' \overline{\partial} \!\cdot\! \overline{\partial}{}'
	-
	(D\!-\!2) H^2aa' \overline{\nabla}{} \!\cdot\! \overline{\nabla}{}'
	\Big]
	\frac{ \Gamma\big(\tfrac{D-2}{2}\big) \, \mu^{D-4} \, i \delta^D(x\!-\!x') }
		{ 16 \pi^{ \frac{D}{2} } (D\!-\!3)(D\!-\!4) }
\nonumber \\
&
 	+
	\kappa^2 \Big[
    2 (m^2\!-\!H^2) aa' 
        \overline{\partial} \!\cdot\! \overline{\partial}{}'
    +
    \overline{\partial} \!\cdot\! \overline{\partial}{}'
        \big( Ha \partial_0 \!+\! Ha' \partial_0' \big)
	-
	H^2 ( a^2 \!+\! a'^2) 
    \overline{\nabla} \!\cdot\! \overline{\nabla}{}'
	\Big]
	\frac{ \partial \!\cdot\! \partial' }{128 \pi^4}
	\bigg( \frac{ \ln(\mu^2\Delta x^2) }{ \Delta x^2 } \bigg)
\nonumber \\
&
	-
	\frac{\kappa^2H^4 (aa')^2}{ 64 \pi^4y}
	\Big[ 
    ( \overline{\partial} \!\cdot\! \overline{\partial}{}' ) 
    ( \partial \!\cdot\! \partial{}')
	+
	( \overline{\partial} \!\cdot\! \partial{}' ) (\overline{\partial}{}' \!\cdot\! \partial)
	-
	( \overline{\partial}{} \!\cdot\! \partial ) ( \overline{\partial}{}' \!\cdot\! \partial{}' )
	\Big]
	\mathcal{F}_m^1
	\, ,
\label{Contracted2I}
\\
\MoveEqLeft[2]
- \kappa^2
\Theta_{I\!I}^{\mu\nu\rho\sigma} (aa')^{D-2} \mathcal{A}_{I\!I} 
	\overline{\partial}{}_\mu\partial{}_\nu \overline{\partial}{}'_\rho\partial{}'_\sigma i \Delta_m
	\ \overset{D\to4}{\longsim} \ 
	\kappa^2 
	H^2aa'
	\Big[ -3(D\!-\!2)(D\!-\!3)\overline{\nabla} \!\cdot\! \overline{\nabla}{}'
\nonumber \\
&	\hspace{1cm}
	- (D\!-\!1)(D\!-\!5)  \overline{\partial}{} \!\cdot\! \overline{\partial}{}' \Big]
	\bigg[
	\frac{ \Gamma\big( \frac{D-2}{2} \big) \, \mu^{D-4} \, i \delta^D(x\!-\!x') }
		{ 16 \pi^{ \frac{D}{2} } (D\!-\!3) (D\!-\!4) }
	+
	\frac{ \partial \!\cdot\! \partial' }{ 128 \pi^4 } \bigg(
		\frac{ \ln(\mu^2 \Delta x^2) }{ \Delta x^2 }
		\bigg)
	\bigg]
\nonumber \\
&
	+
	\frac{\kappa^2 H^4 (aa')^2}{128\pi^4}
	\big[ 2 \!-\! 2\gamma_{\rm \scr E} \!+\! \ln(aa')\big]
	\Big[
	3 ( \overline{\partial}{} \!\cdot\! \partial ) ( \overline{\partial}{}' \!\cdot\! \partial{}' )
	-
	2 ( \overline{\nabla}{} \!\cdot\! \nabla ) ( \overline{\partial}{}' \!\cdot\! \partial{}' )
	-
	2 ( \overline{\partial}{} \!\cdot\! \partial ) ( \overline{\nabla}{}' \!\cdot\! \nabla{}' )
	\Big]
	\Big( \frac{4}{y} \!+\! \mathcal{F}_m^0 \Big)
\nonumber \\
&
	-
	\frac{ \kappa^2 H^4(aa')^2 }{ 32 \pi^4 }
	\Big[
	3 ( \overline{\partial}{} \!\cdot\! \partial ) ( \overline{\partial}{}' \!\cdot\! \partial{}' )
	-
	2 ( \overline{\nabla}{} \!\cdot\! \nabla ) ( \overline{\partial}{}' \!\cdot\! \partial{}' )
	-
	2 ( \overline{\partial}{} \!\cdot\! \partial ) ( \overline{\nabla}{}' \!\cdot\! \nabla{}' )
	\Big]
	\frac{ 2 \!+\! \ln(y) }{ y }
\nonumber \\
&
	-
	\frac{\kappa^2 H^4 (aa')^2}{128\pi^4}
	\ln(y)
	\Big[
	3 ( \overline{\partial}{} \!\cdot\! \partial ) ( \overline{\partial}{}' \!\cdot\! \partial{}' )
	-
	2 ( \overline{\nabla}{} \!\cdot\! \nabla ) ( \overline{\partial}{}' \!\cdot\! \partial{}' )
	-
	2 ( \overline{\partial}{} \!\cdot\! \partial ) ( \overline{\nabla}{}' \!\cdot\! \nabla{}' )
    -
    (\overline{\nabla} \!\cdot\! \overline{\nabla}{}') 
        ( \nabla \!\cdot\! \nabla{}' )
\nonumber \\
&	\hspace{1cm}
    +
    (\overline{\nabla} \!\cdot\! \nabla) 
        ( \overline{\nabla}{}' \!\cdot\! \nabla{}' )
	\Big]
	\mathcal{F}_m^0
	+
	\frac{ \kappa^2 H^3aa'}{ 64 \pi^4 }
	\Big[ 2 \big(a' \partial_0
		      \!+\! a \partial{}'_0 \big)
		+ 3aa'H \Big]
		\overline{\nabla} \!\cdot\! \overline{\nabla}{}' 
        \frac{1}{\Delta x^2}
	\, ,
\label{Contracted2II}
\\
\MoveEqLeft[2]
-\kappa^2
\Theta_{I\!I\!I}^{\mu\nu\rho\sigma} (aa')^{D-2} \mathcal{A}_{I\!I\!I}
	\overline{\partial}{}_\mu\partial{}_\nu \overline{\partial}{}'_\rho\partial{}'_\sigma i \Delta_m
	\ \overset{D\to4}{\longsim} \ 
		- \kappa^2 H^2aa' \big(
		3 \overline{\partial}{} \!\cdot\! \overline{\partial}{}'
		- 
		2 \overline{\nabla}{} \!\cdot\! \overline{\nabla}{}'
		\big)
		\frac{i \delta^4(x\!-\!x') }{ 32 \pi^2 }
	\, .
\label{Contracted2III}
\end{align}
%

\subsubsection{Consolidation and renormalization}
\label{subsubsec: Consolidation and renormalization}

We now combine the different parts of each contribution
$I$--$I\!I\!I$ and absorb their ultraviolet divergences,
together with the chosen finite local terms, into
the counterterms~(\ref{SelfMassCtm}). We renormalize the
three contributions separately, so that the 3-vertex
contributions to the counterterm coefficients decompose as
\begin{equation}
\alpha_n^{\rm 3pt} = \alpha_n^{{\rm 3pt}, I}
    + \alpha_n^{{\rm 3pt}, I\!I}
    + \alpha_n^{{\rm 3pt}, I\!I\!I}
    \, ,
\qquad\quad
\beta_n^{\rm 3pt} = \beta_n^{{\rm 3pt}, I}
    + \beta_n^{{\rm 3pt}, I\!I}
    + \beta_n^{{\rm 3pt}, I\!I\!I}
    \, .
\end{equation}
In addition, we simplify the finite nonlocal terms by 
extracting and rearranging the derivatives acting on them.
The identities required for these manipulations are
collected in Appendix~\ref{subapp: Identities used with consolidation}.

\paragraph{Contribution $I$.}
Combining Eqs.~(\ref{Contracted0I}),
(\ref{Contracted1I}), and~(\ref{Contracted2I}) gives the
complete 3-vertex contribution associated with part $I$ of
the graviton propagator:
\begin{align}
\MoveEqLeft[2]
\big[ -i\mathcal{M}^2_{\rm 3pt} (x;x') \big]_I 
	\ \overset{D\to4}{\longsim} \
	\kappa^2 \bigg[
	4m^2 aa' \overline{\partial}{} \!\cdot\! \overline{\partial}{}'
	-
	\frac{(D\!+\!2)}{2} H^2 aa' \overline{\partial} \!\cdot\! \overline{\partial}'
	-
	(D\!-\!2) H^2aa' \overline{\nabla}{} \!\cdot\! \overline{\nabla}{}'
	+
	D m^4 (aa')^2
\nonumber \\
&
	-
	7 (D\!-\!4) m^2 H^2 (aa' )^2
	\bigg]
	\frac{ \Gamma\big(\tfrac{D-2}{2}\big) \mu^{D-4} i \delta^D(x\!-\!x') }
		{ 16 \pi^{ \frac{D}{2} } (D\!-\!3)(D\!-\!4) }
 	+
	\kappa^2
	\bigg[
    2 m^2 aa'
		\overline{\partial} \!\cdot\! \overline{\partial}{}'
	- H^2 (a^2 \!+\! a'^2)
        \overline{\nabla} \!\cdot\! \overline{\nabla}{}'
\nonumber \\
&
    +
    \overline{\partial} \!\cdot\! \overline{\partial}{}'
    \big( Ha \partial_0 \!+\! Ha' \partial_0' 
        \!-\! 2 H^2 aa' \big)
    -
    \big( a'^4 \overline{\mathcal{D}} + a^4 \overline{\mathcal{D}}{}'
		\big) \frac{m^2}{(aa')^2}
	+
    4 m^4 (aa')^2
	\bigg]
	\frac{ \partial \!\cdot\! \partial' }{128 \pi^4}
	\bigg( \frac{ \ln(\mu^2\Delta x^2) }{ \Delta x^2 } \bigg)
\nonumber \\
&
	-
	\frac{\kappa^2H^4 (aa')^2}{ 64 \pi^4y}
	\Big[ 
    ( \overline{\partial} \!\cdot\! \overline{\partial}{}' ) 
    ( \partial \!\cdot\! \partial{}')
	+
	( \overline{\partial} \!\cdot\! \partial{}' ) (\overline{\partial}{}' \!\cdot\! \partial)
	-
	( \overline{\partial}{} \!\cdot\! \partial ) ( \overline{\partial}{}' \!\cdot\! \partial{}' )
	\Big]
	\mathcal{F}_m^1
	+
	\frac{\kappa^2 m^4 H^4(aa')^4}{32 \pi^4 y}
        \mathcal{F}_m^0
\nonumber \\
&
	+
	\frac{\kappa^2 m^2H^4(aa')^2}{ 64 \pi^4 y}
	\big(
	a'{}^2 \, \overline{\partial}{} \!\cdot\! \partial 
	+
	a^2 \, \overline{\partial}{}' \!\cdot\! \partial{}'
	\big) \mathcal{F}_m^1
	+
	\frac{ \kappa^2m^2 (m^2 \!-\! 2H^2) H^2 (aa')^3 }
        { 16 \pi^4 \Delta x^2}
	\, .
\end{align}
The nonlocal terms in the last two lines are simplified
using Eqs.~(\ref{contIidentity1})--(\ref{contIidentity3}).
The ultraviolet divergences, together with the chosen finite 
local terms, are absorbed by the flat-space counterterms
\begin{equation}
\alpha^{{\rm 3pt},I}_{1}
    =
    - \frac{4 \widetilde{\mu}{}^{D-4}}{D\!-\!4} \, ,
\qquad\quad
\alpha^{{\rm 3pt},I}_{2}
    =
    \frac{(D\!-\!8) \widetilde{\mu}{}^{D-4}}{D\!-\!4} \, ,
\end{equation}
and the curvature-dependent counterterms
\begin{equation}
\beta^{{\rm 3pt},I}_{1}
    =
    \frac{(D\!+\!2) \widetilde{\mu}{}^{D-4}}{2(D\!-\!4)}
    \, ,
\qquad\quad
\beta^{{\rm 3pt},I}_{2}
    =
    3 \widetilde{\mu}{}^{D-4} \, ,
\qquad\quad
\beta^{{\rm 3pt},I}_{3}
    =
    \frac{(D\!-\!2) \widetilde{\mu}{}^{D-4}}{D-4} \, ,
\end{equation}
where $\widetilde{\mu}^{D-4}$ is defined in
(\ref{MuTildeDef}). With this choice of counterterms,
the renormalized contribution is
\begin{align}
\MoveEqLeft[2]
\big[ -i\mathcal{M}^2_{\rm 3pt} (x;x') \big]_I^{\rm ren}
	=
    \frac{\kappa^2 }{32 \pi^2}
    \Big[
	(3H^2 \!-\! 4 m^2)
		\overline{\partial} \!\cdot\! \overline{\partial}{}'
	+
	2 H^2 \overline{\nabla} \!\cdot\! \overline{\nabla}{}'
	-
    4 m^4 aa'
	\Big]
    aa' \ln(aa') 
	i \delta^4(x\!-\!x')
\nonumber \\
&
 	+
	\kappa^2
	\bigg[
    2 m^2 aa'
		\overline{\partial} \!\cdot\! \overline{\partial}{}'
	- H^2 (a^2 \!+\! a'^2)
        \overline{\nabla} \!\cdot\! \overline{\nabla}{}'
    +
    H\overline{\partial} \!\cdot\! \overline{\partial}{}'
    \big( a \partial_0 \!+\! a' \partial_0' 
        \!-\! 2 H aa' \big)
    -
    \big( a'^4 \overline{\mathcal{D}} 
        \!+\! a^4 \overline{\mathcal{D}}{}'
		\big) \frac{m^2}{(aa')^2}
\nonumber \\
&   \hspace{0.8cm}
	+
    4 m^4 (aa')^2
	\bigg]
	\frac{ \partial \!\cdot\! \partial' }{128 \pi^4}
	\bigg( \frac{ \ln(\mu^2\Delta x^2) }{ \Delta x^2 } \bigg)
	-
	\frac{\kappa^2(m^2 \!-\! 2H^2)H aa'}{ 64 \pi^4}
	\overline{\partial} \!\cdot\! \overline{\partial}{}'
    \big( a\partial'_0 \!+\! a'\partial_0 \big)
	\frac{1}{\Delta x^2}
\nonumber \\
&
	-
	\frac{\kappa^2H^4}{ 64 \pi^4}
    \bigg[
    H(aa')^2 
    \Big(
    \big( a\partial'_0 \!+\! a'\partial_0 \big)
    \overline{\partial} \!\cdot\! \overline{\partial}{}'
    +
    ( a \!+\! a' )
    ( \overline{\partial}{}'_0 \!+\! \overline{\partial}_0 )
    \overline{\nabla} \!\cdot\! \overline{\nabla}{}'
    \Big)
    +
    m^2\big(
        a'^4 \overline{\mathcal{D}}
        +
        a^4 \overline{\mathcal{D}}{}'
        \big)
    \bigg]
	I \Big[
        \frac{1}{y}\frac{\partial \mathcal{F}_m^0}{\partial y}
        \Big]
\nonumber \\
&
	-
	\frac{\kappa^2H^5(aa')^2(a\!+\!a')}{ 128 \pi^4}
    \bigg[
    \overline{\partial} \!\cdot\! \overline{\partial}{}'
    \Big(
    \overline{\partial}{}'_0 \!+\! \overline{\partial}_0
    \!-\! H(a\!+\!a')
    \Big)
    +
    2 \overline{\nabla} \!\cdot\! \overline{\nabla}{}'
	( \overline{\partial}{}'_0 \!+\! \overline{\partial}_0 )
    \bigg]
    \frac{\partial\mathcal{F}_m^0}{\partial y}
\nonumber \\
&
	+
    \frac{\kappa^2 m^2H^4(aa')^2}{128\pi^4}
    \Big[
	( a^2 \!+\! a'^2 )
	\overline{\partial} \!\cdot\! \overline{\partial}{}'
	+
	4 m^2 (aa')^2
    \Big]
    \frac{\mathcal{F}_m^0}{y}
	\, ,
\label{Irenormalized}
\end{align}
where we introduce the notation for primitive functions
\begin{equation}
I\big[f\big](y)  \equiv \int^y \! ds \, f(s) \, .
\end{equation}

In the flat-space limit,~$H\!\to\!0$, 
this contribution reproduces the 
full Minkowski-space result~(\ref{FlatSelfMass1}), up to the 
finite local counterterm contributions, which are not fixed by 
renormalization.
This follows by using the flat-space limits:
\begin{equation}
y \ \overset{H \to 0}{\longsim} \ H^2 \Delta x^2 \, ,
\qquad
\frac{\partial}{\partial y} \ \overset{H \to 0}{\longsim} \
    \frac{1}{H^2} \frac{\partial}{\partial \Delta x^2}
    \, ,
\qquad
I[...] \ \overset{H \to 0}{\longsim} \
    H^2 \mathcal{I}[...]
    \, ,
\qquad
\mathcal{F}_m^0
    \ \overset{H \to 0}{\longsim} \
    \frac{m^2}{H^2} f_m^0
    \, .
\label{FlatLimits}
\end{equation}

\paragraph{Contribution $I\!I$.}
Combining Eqs.~(\ref{Contracted0II}),
(\ref{Contracted1II}), and~(\ref{Contracted2II}) gives the
complete 3-vertex contribution associated with part
$I\!I$ of the graviton propagator:
\begin{align}
\MoveEqLeft[2]
- i \mathcal{M}^2_{{\rm 3pt},I\!I}(x;x')
	=
    \kappa^2 
	H^2aa'
	\Big[ -3(D\!-\!2)(D\!-\!3)\overline{\nabla} \!\cdot\! \overline{\nabla}{}'
	-
    (D\!-\!1)(D\!-\!5)  \overline{\partial}{} \!\cdot\! \overline{\partial}{}' \Big]
\nonumber \\
&   \hspace{0.5cm}
    \times\!
    \bigg[
	\frac{ \Gamma\big( \frac{D-2}{2} \big) \, \mu^{D-4} \, i \delta^D(x\!-\!x') }
		{ 16 \pi^{ \frac{D}{2} } (D\!-\!3) (D\!-\!4) }
	+
	\frac{ \partial \!\cdot\! \partial' }{ 128 \pi^4 } \bigg(
		\frac{ \ln(\mu^2 \Delta x^2) }{ \Delta x^2 }
		\bigg)
	\bigg]
	+
	\frac{\kappa^2 H^4 (aa')^2}{128\pi^4}
	\big[ 2 \!-\! 2\gamma_{\rm \scr E} \!+\! \ln(aa')\big]
\nonumber \\
&   \hspace{0.5cm}
    \times\! \Big[
	3 \bigl( \overline{\partial}{} \!\cdot\! \partial
        \!+\! m^2 a^2 \bigr) 
        \bigl( \overline{\partial}{}' \!\cdot\! \partial{}' 
            \!+\! m^2 a'^2 \bigr)
	-
	2 \overline{\nabla}{} \!\cdot\! \overline{\nabla}{}'
        \bigl( \overline{\partial}{} \!\cdot\! \partial
            \!+\! m^2 a^2
            \!+\!
            \overline{\partial}{}' \!\cdot\! \partial{}'
            \!+\! m^2 a'^2 
            \bigr)
	\Big]
	\Big( \frac{4}{y} \!+\! \mathcal{F}_m^0 \Big)
\nonumber \\
&
	-
	\frac{ \kappa^2 H^4(aa')^2 }{ 32 \pi^4 }
	\Bigl[
	3 \overline{\partial}{} \!\cdot\! \partial
        \bigl( \overline{\partial}{}' \!\cdot\! \partial{}'
            \!+\! m^2 a'^2 \bigr)
    +
    3 \overline{\partial}{}' \!\cdot\! \partial{}'
    \bigl( \overline{\partial}{} \!\cdot\! \partial
        \!+\! m^2 a^2 \bigr)
    -
	2 \overline{\nabla}{} \!\cdot\! \overline{\nabla}{}'
    \bigl( 2 \overline{\partial}{} \!\cdot\! \partial
        \!+\! m^2 a^2
\nonumber \\
&   \hspace{1cm}
        \!+\!
        2 \overline{\partial}{}' \!\cdot\! \partial{}' 
        \!+\! m^2 a'^2
        \bigr)
    \Bigr] \frac{1}{y}
	+
	\frac{ \kappa^2 H^3aa'}{ 64 \pi^4 }
	\Big[ 2 a' \partial_0
		      + 2a \partial{}'_0
		+ 3aa'H \Big]
		\overline{\nabla} \!\cdot\! \overline{\nabla}{}' 
        \frac{1}{\Delta x^2}
\nonumber \\
&
	-
	\frac{ \kappa^2 H^4(aa')^2 }{ 32 \pi^4 }
	\Bigl[
	3 \bigl( \overline{\partial}{} \!\cdot\! \partial
        \!+\! m^2 a^2 \bigr) 
        \bigl( \overline{\partial}{}' \!\cdot\! \partial{}' 
            \!+\! m^2 a'^2 \bigr)
	-
	2 \overline{\nabla}{} \!\cdot\! \overline{\nabla}{}'
    \bigl( \overline{\partial}{}' \!\cdot\! \partial{}'
        \!+\! m^2 a'^2
        \!+\!
        \overline{\partial}{} \!\cdot\! \partial
        \!+\! m^2 a^2\bigr)
    \Bigr]
    \frac{\ln(y)}{y}
\nonumber \\
&
	-
	\frac{\kappa^2 H^4 (aa')^2}{128\pi^4}
	\ln(y)
	\Big[
	3 \bigl( \overline{\partial}{} \!\cdot\! \partial 
        \!+\! m^2 a^2 \bigr)
        \bigl( \overline{\partial}{}' \!\cdot\! \partial{}' 
            \!+\! m^2 a'^2 \bigr)
	-
	2 \overline{\nabla}{} \!\cdot\! \nabla
        \bigl( \overline{\partial}{}' \!\cdot\! \partial{}'
            \!+\! m^2 a'^2 \bigr)
\nonumber \\
&   \hspace{1cm}
	-
	2 \overline{\nabla}{}' \!\cdot\! \nabla{}'
    \bigl( \overline{\partial}{} \!\cdot\! \partial 
        \!+\! m^2 a^2 \bigr)
    -
    (\overline{\nabla} \!\cdot\! \overline{\nabla}{}') 
        ( \nabla \!\cdot\! \nabla{}' )
    +
    (\overline{\nabla} \!\cdot\! \nabla) 
        ( \overline{\nabla}{}' \!\cdot\! \nabla{}' )
	\Big]
	\mathcal{F}_m^0
    \, .
\end{align}
The nonlocal terms, comprising most of the expression above,
are simplified using the nine identities~(\ref{contIIidentity1})--(\ref{contIIidentity9}).
This contribution requires no flat-space counterterms,
$\alpha^{{\rm 3pt},I\!I}_{1} \!=\! \alpha^{{\rm 3pt},I\!I}_{2}
    \!=\! 0$.
But its local divergences, and some of the finite local parts, 
are instead absorbed by the curvature-dependent counterterms,
\begin{equation}
\beta^{{\rm 3pt},I\!I}_{1} 
    =
    - \frac{ (D\!-\!1)(D\!-\!3)^3 
        \widetilde{\mu}{}^{D-4}}{D\!-\!4}
    \, ,
\qquad\
\beta^{{\rm 3pt},I\!I}_{2}
    =
    - 9\widetilde{\mu}{}^{D-4}
    \, ,
\qquad\
\beta^{{\rm 3pt},I\!I}_{3}
    =
    \frac{ (D\!+\!2)(D\!-\!3)^4
        \widetilde{\mu}{}^{D-4}}{D\!-\!4}
    \, .
\end{equation}

With this choice of counterterms, the renormalized
contribution $I\!I$  is
\begin{align}
\MoveEqLeft[2]
\bigl[ -i\mathcal{M}^2_{\rm 3pt} (x;x') \bigr]_{I\!I}^{\rm ren}
	=
    \frac{3 \kappa^2 H^2aa'}{32 \pi^2}
	\bigl(
    2 \overline{\nabla} \!\cdot\! \overline{\nabla}{}'
    \!-\!
    \overline{\partial}{} \!\cdot\! \overline{\partial}{}'
    \bigr)
    \bigg[
    \ln(aa') i \delta^4(x\!-\!x')
	-
	\frac{ \partial \!\cdot\! \partial' }{ 4 \pi^2 }
        \bigg(
		\frac{ \ln(\mu^2 \Delta x^2) }{ \Delta x^2 }
		\bigg)
	\bigg]
\nonumber \\
&
	+
	\frac{3 \kappa^2 H^4}{128\pi^4}
    \big( \overline{\mathcal{D}} \!-\! m^2 a^4 \big)
    \big( \overline{\mathcal{D}}{}' \!-\! m^2 a'^4 \big)
    \biggl(
    \big[ 2 \!-\! 2\gamma_{\scr \rm E} \!-\! 
        \ln(H^2 \Delta x^2) \big]
    \Big( \frac{4}{y} \!+\! \mathcal{F}_m^0 \Big)
    +
    \frac{3}{2} I \Big[ \frac{\mathcal{F}_m^0}{y} \Big]
    \biggr)
\nonumber \\
&
	+
	\frac{\kappa^2 H^4}{64\pi^4}
    \overline{\nabla}{} \!\cdot\! \overline{\nabla}{}'
	\Big[
		a'^2 \big( \overline{\mathcal{D}} \!-\! m^2a^4 \big)
        +
		a^2 \big( \overline{\mathcal{D}}{}' \!-\! m^2a'^4 \big)
	\Big]
    \biggl(
    \big[ 2 \!-\! 2\gamma_{\scr \rm E} \!-\! 
        \ln(H^2 \Delta x^2) \big] 
    \Big( \frac{4}{y} \!+\!  \mathcal{F}_m^0 \Big)
    +
    I \Big[ \frac{\mathcal{F}_m^0}{y} \Big]
    \biggr)
\nonumber \\
&
	+
	\frac{3 \kappa^2 H^4}{128\pi^4}
    \Big[
    a'^4 \big( \overline{\mathcal{D}} \!-\! m^2 a^4 \big)
    +
    a^4 \big( \overline{\mathcal{D}}{}' \!-\! m^2 a'{}^4 \big)
    \Big]
    \bigg(
	\frac{3m^2}{4} I \Big[ \frac{\mathcal{F}_m^0}{y} \Big]
    \!+\!
    H^2 I \Big[ \frac{\mathcal{F}_m^0}{y^2} \Big]
    \bigg)
\nonumber \\
&
	+
	\frac{\kappa^2 H^4}{64\pi^4}
    \overline{\nabla}{} \!\cdot\! \overline{\nabla}{}'
	\Big[
		a'^2 \big( \overline{\mathcal{D}}
            \!-\! Ha^3 \overline{\partial}_0 \big)
        +
		a^2 \big( \overline{\mathcal{D}}{}'
            \!-\! Ha'^3 \overline{\partial}{}'_0 \big)
	\Big]
    I^2 \Big[ \frac{1}{y}
    \frac{\partial \mathcal{F}_m^0}{\partial y} \Big]
	+
	\frac{3 \kappa^2 m^2 H^6 (aa')^4 }{64\pi^4 }
    I \Big[ \frac{\mathcal{F}_m^0}{y^2} \Big]
\nonumber \\
&
	-
	\frac{\kappa^2 H^6 (aa')^2 }{64\pi^4}
    \overline{\nabla} \!\cdot\! \overline{\nabla}{}'
    \biggl(
    6 aa' I \Big[ \frac{1}{y} 
        \frac{\partial \mathcal{F}_m^0}{\partial y} \Big]
        +
        \big( a \overline{\partial}_0
        \!+\!
        a' \overline{\partial}{}'_0
        \big)
    \mathcal{F}_m^0
    \biggr)
	+
	\frac{9 \kappa^2 (m^2 \!-\! 2H^2) H^4 (aa')^3}
        {32\pi^4 \Delta x^2}
\nonumber \\
&
	+
	\frac{ \kappa^2 H^2 aa'}{ 64 \pi^4 }
	\Big[ 2 Ha \partial{}_0 + 2 Ha' \partial_0'
		+ 3H^2aa' +
        4 (m^2 \!-\! 2H^2) (a^2 \!+\! a'^2)
        \Big]
		\overline{\nabla} \!\cdot\! \overline{\nabla}{}' 
        \frac{1}{\Delta x^2}
    \, .
\label{IIrenormalized}
\end{align}
Using the flat-space limits in Eq.~(\ref{FlatLimits}), 
this contribution vanishes in the Minkowski limit.

\paragraph{Contribution $I\!I\!I$.}
The final contribution to the self-mass is local and finite:
\begin{equation}
\big[ - i \mathcal{M}^2_{\rm 3pt} (x;x') \big]_{I\!I\!I}
	=
	-
	\frac{\kappa^2 H^2aa'}{ 32 \pi^2  }
	\big(
	3 \overline{\partial}{} \!\cdot\! \overline{\partial}{}'
	- 
	2 \overline{\nabla}{} \!\cdot\! \overline{\nabla}{}'
	\big)
	i \delta^4(x\!-\!x')
	\, .
\end{equation}
It can be absorbed entirely into the finite parts of the
curvature-dependent counterterms,
\begin{equation}
\alpha^{{\rm 3pt},I\!I\!I}_{1}
    =
    \alpha^{{\rm 3pt},I\!I\!I}_{2} 
    =
    \beta^{{\rm 3pt},I\!I\!I}_{2}
    = 0 \, ,
\qquad\quad
\beta^{{\rm 3pt},I\!I\!I}_{1}
    =
    \frac{3\widetilde{\mu}{}^{D-4}}{2} \, ,
\qquad\quad
\beta^{{\rm 3pt},I\!I\!I}_{3} 
    =
    -2 \widetilde{\mu}{}^{D-4} \, ,
\end{equation}
which leaves the renormalized contribution vanishing,
\begin{equation}
\big[ - i \mathcal{M}^2_{\rm 3pt} (x;x')
    \big]_{I\!I\!I}^{\rm ren}
	=
	0
	\, .
\end{equation}
%

\section{Discussion}
\label{sec: Discussion}

In this work we presented a dimensionally regulated and
renormalized computation of the one-graviton-loop
self-mass of a massive, minimally coupled scalar field in
de Sitter space. The total self-mass is the sum of the
three contributions~(\ref{4ptRenormalized}),
(\ref{Irenormalized}), and~(\ref{IIrenormalized}) obtained
in Sec.~\ref{sec: Evaluating one-loop diagrams in de Sitter
space}, supplemented by the finite local contributions from
the counterterms~(\ref{SelfMassCtm}). Since combining these
terms produces no further useful simplifications or
cancellations, we do not display the resulting lengthy
expression in a single formula. We have checked that the
final result correctly reproduces the flat-space result,
which was computed independently in
Sec.~\ref{sec: Self-mass in Minkowski space}.

Our primary motivation for computing the massive scalar
self-mass is its role in constructing a gauge-independent
effective field equation for a massless, minimally coupled scalar in de Sitter. 
Recent work has shown that
one-loop corrections to the external mode functions of the
massive source and observer fields are essential for
canceling the graviton gauge dependence of this equation
\cite{Glavan:2026pug}. The self-mass obtained here provides
the ingredient required to determine those corrections in
de Sitter space. The relevant part of the self-mass for this purpose is the local terms containing secular logarithms --- the first terms in~(\ref{4ptRenormalized}), (\ref{Irenormalized}), and~(\ref{IIrenormalized}), which can be written in the following form
\begin{equation}
-i \mathcal{M}^2_{\rm ren}(x;x')
    \supset
    \frac{\kappa^2}{8\pi^2}
    \bigg[
    (m^2 \!-\! 3H^2) \Big[
		\big( \mathcal{D} \!-\! m^2a^4 \big)
		+
		\big( \mathcal{D}{}' \!-\! m^2a'^4 \big)
		\Big]
	+
	4H^2 a^2 \nabla^2
    \bigg]
    \ln(a) i \delta^4(x\!-\!x')
    \, .
\label{LogCor}
\end{equation}
These arise from the UV corrections and are always 
accompanied by the logarithm of the renormalization 
scale~$\mu$, divided by the appropriate physical 
scale.\footnote{This follows from using the ~$D\!=\!4$
specialization of the identity
(\ref{FlatPropEOM}) in Eqs.~(\ref{Irenormalized}) 
and~(\ref{IIrenormalized}).}
Combined with 
the source and observer contributions, the external 
wavefunction corrections generated by~(\ref{LogCor}) have
allowed us to infer the effective gauge-independent self-mass
\cite{Glavan:2021adm,Glavan:2024elz,Glavan:2026bfw}, and to complete
the one-loop gauge-independent correction to the 
tree-level potential generated by
a scalar point source in de Sitter space~\cite{Burko:2002ge,Akhmedov:2010ah,Glavan:2019yfc}.\footnote{The same superhorizon scalar potential is
generated by any localized, static, and spherically
symmetric source coupled linearly to the scalar
\cite{Glavan:2026jxe}.}

No local terms containing secular logarithms 
of the form in~(\ref{LogCor}) arise from 
the remaining contributions in~(\ref{Irenormalized}) 
and~(\ref{IIrenormalized}). This is, however, not immediately
obvious, but for most terms it can be established by 
considering separately terms with different derivative 
structures and examining their most singular terms using 
the power-series representation in~(\ref{FpowerSeries}). 
That is true for all terms except for the one in the
fourth line in~(\ref{IIrenormalized}): one third of the 
first term in the brackets should be rewritten using the
identity given in~(\ref{LastIdentity}) to establish that 
no contributions such as the ones in~(\ref{LogCor})
appear from the IR.

Because the mass has been kept arbitrary, the result also
applies beyond this immediate objective. For example, it
can be used to study one-graviton-loop corrections,
in the simple gauge~\cite{Tsamis:1992xa,Woodard:2004ut}, to
one- and two-point functions of light or heavy massive scalars. This
opens the way to determining how the infrared enhancements
known for massless fields are modified at nonzero mass, how
they depend on $m/H$, and how they connect to the
heavy-field regime.

The self-mass admits several equivalent representations,
whose equivalence is not always manifest. We have chosen a
form in which the ultraviolet structure is explicit and the
renormalization scale $\mu$ is cleanly separated from the
physical scales. This organization also makes manifest that
the renormalization scale~$\mu$ always appears together
with a factor of the scale factor~$a$. Such terms,
originating from ultraviolet effects, are expected to
generate enhanced secular corrections to the mode functions
of very massive scalars, which are amenable to resummation
using renormalization-group techniques. More generally,
secular corrections in de Sitter space can originate from
both ultraviolet and infrared effects. The latter require
additional resummation methods adapted to long-wavelength
dynamics, such as stochastic inflation
\cite{Starobinsky:1986fx,Starobinsky:1994bd}. In situations
where both types of secular enhancement are present, both
resummation methods are therefore    required
\cite{Miao:2021gic,Woodard:2023rqo,Glavan:2023lvw,
Litos:2023nvj,Glavan:2023tet,Miao:2024nsz,
Foraci:2024vng,Foraci:2024cwi}.

Other representations remain useful for comparison and for
specific applications. The identities collected in
Appendix~\ref{app: Additional useful identities} permit the
result to be rearranged into such forms, as illustrated for
the Minkowski-space self-mass in
Sec.~\ref{sec: Self-mass in Minkowski space}. They were
also instrumental in comparing the present calculation with
an alternative organization in which integration by parts
and vertex identities are applied before the ultraviolet
divergences are isolated. The most useful representation of
the self-mass therefore depends on the particular
application and the question being addressed. The
renormalized result derived here provides a basis for
pursuing these questions for massive scalar fields in de
Sitter space.

\section*{Acknowledgments}
\addcontentsline{toc}{section}{\protect\numberline{}Acknowledgments} 

The authors acknowledge a generous travel support by the Delta ITP consortium, a program of
the Netherlands Organisation for Scientific Research (NWO) that is funded by the Dutch
Ministry of Education, Culture and Science (OCW) --- NWO project number 24.001.027.
DG was supported by the Czech Science Foundation (GA\v{C}R) grant 24-13079S. 
SPM was partially supported by Taiwan NSTC grants 113-2112-M-006-013 and 114-2112-M-006-020.
TP is supported by the NWA ORC 2023 consortium grant: Cosmic emergence: from abstract simplicity to complex diversity (Kosmische
emergentie: van abstracte eenvoud naar complexe diversiteit).
DG and TP are funded by The Magnetic Universe NWO grant OCENW.XL.23.147.
RPW was partially supported by NSF grant PHY-2207514 and by the Institute for 
Fundamental Theory at the University of Florida.
DG is grateful to Alberto Gemma for his gracious help
with the logistics during this project.

\appendix

\section{Identities for localizing divergences}
\label{app: Localizing divergences}

The following identities for extracting derivatives 
from products are easily proved:
\begin{subequations}
\begin{align}
\frac{1}{\Delta x^{2D-4}}
	={}&
	- \frac{\partial \!\cdot\! \partial'}{2(D\!-\!3)(D\!-\!4)} \frac{1}{\Delta x^{2D-6}}
	\, ,
\\
\frac{1}{\Delta x^{D-2}} \partial_\mu \frac{1}{\Delta x^{D-2}}
	={}&
	- \frac{ \partial_\mu (\partial \!\cdot\! \partial') }
			{4(D\!-\!3)(D\!-\!4)} \frac{1}{\Delta x^{2D-6}}
	\, ,
\\
\frac{1}{\Delta x^{D-2}} \partial_\mu \frac{1}{\Delta x^{D-4}}
	={}&
	\frac{(D\!-\!4) \partial_\mu }{2(D\!-\!3)} \frac{1}{\Delta x^{2D-6}}
	\, ,
\\
\frac{1}{\Delta x^{D-4}} \partial_\mu \frac{1}{\Delta x^{D-2}}
	={}&
	\frac{(D\!-\!2) \partial_\mu }{2(D\!-\!3)} \frac{1}{\Delta x^{2D-6}}
	\, ,
\\
\frac{1}{\Delta x^{D-2}} \partial_\mu \partial'_\nu \frac{1}{\Delta x^{D-2}}
	={}&
	-
	\frac{ \big( D \partial_\mu \partial{}'_\nu - \eta_{\mu\nu} \partial \!\cdot\! \partial' \big) 	
		\partial \!\cdot\! \partial'}{8(D\!-\!1)(D\!-\!3)(D\!-\!4)} 
	\frac{1}{\Delta x^{2D-6}}
	\, ,
\\
\frac{1}{\Delta x^{D-2}} \partial_\mu \partial'_\nu \frac{1}{\Delta x^{D-4}}
	={}&
	\frac{(D\!-\!4) \partial_\mu \partial{}'_\nu - \eta_{\mu\nu} \partial \!\cdot\! \partial' }
		{4(D\!-\!3)}
	 	\frac{1}{\Delta x^{2D-6}}
	 \, ,
\\
\frac{1}{\Delta x^{D-4}} \partial_\mu \partial'_\nu \frac{1}{\Delta x^{D-2}}
	={}&
	\frac{D \partial_\mu \partial{}'_\nu - \eta_{\mu\nu} \partial \!\cdot\! \partial' }{4(D\!-\!3)}  	
		\frac{1}{\Delta x^{2D-6}}
	\, .
\end{align}
\end{subequations}
Use the massless propagator equation of motion:
\begin{equation}
\partial \!\cdot\! \partial' \frac{1}{\Delta x^{D-2}}
	=
	- \frac{4\pi^{\frac{D}{2} }  i \delta^D(x\!-\!x') }{ \Gamma\big( \frac{D-2}{2} \big) }
	\, .
\label{FlatPropEOM}
\end{equation}
to derive the identity for localizing non-integrable divergences~\cite{Onemli:2002hr}
\begin{equation}
\partial \!\cdot\! \partial' \frac{1}{\Delta x^{2D-6}}
	\ \overset{D\to4}{\longsim} \ 
	- \frac{4\pi^{\frac{D}{2} }  \mu^{D-4} \, i \delta^D(x\!-\!x') }
		{ \Gamma\big( \frac{D-2}{2} \big) }
	-
	\frac{ (D\!-\!4) \partial \!\cdot\! \partial'}{2} \bigg(
		\frac{ \ln(\mu^2 \Delta x^2) }{ \Delta x^2 }
		\bigg)
		\, .
\end{equation}
We use this to isolate divergences from the expression above:
\begin{subequations}
\begin{align}
&
\frac{1}{\Delta x^{2D-4}}
\ \overset{D\to4}{\longsim} \ 
	\frac{2\pi^{\frac{D}{2} } \mu^{D-4} \, i \delta^D(x\!-\!x') }
		{ (D\!-\!3)(D\!-\!4) \, \Gamma\big( \frac{D-2}{2} \big) }
	+
	\frac{ \partial \!\cdot\! \partial' }{4} 
		\bigg( \frac{ \ln(\mu^2\Delta x^2) }{ \Delta x^2 } \bigg)
	\, ,
\label{1stIDextraction}
\\
&
\frac{1}{\Delta x^{D-2}} \partial_\mu \frac{1}{\Delta x^{D-2}}
\ \overset{D\to4}{\longsim} \ 
	\partial_\mu \bigg[
	\frac{\pi^{\frac{D}{2} }  \mu^{D-4} \, i \delta^D(x\!-\!x') }
		{ (D\!-\!3)(D\!-\!4) \, \Gamma\big( \frac{D-2}{2} \big) }
	+
	\frac{ \partial \!\cdot\! \partial' }{8} \bigg(
		\frac{ \ln(\mu^2 \Delta x^2) }{ \Delta x^2 } \bigg)
	\bigg]
	\, ,
\\
&
\frac{1}{\Delta x^{D-2}} \partial_\mu \frac{1}{\Delta x^{D-4}}
\ \overset{D\to4}{\longsim} \ 
	\frac{(D\!-\!4) }{2} \partial_\mu \frac{1}{\Delta x^2}
	\, ,
\\
&
\bigg[ \frac{1}{\Delta x^{D-4}} - \Big( \frac{H^2aa'}{4} \Big)^{\!\frac{D-4}{2}} \bigg]
	\partial_\mu \frac{1}{\Delta x^{D-2}}
\ \overset{D\to4}{\longsim} \ 
	\frac{(D\!-\!4)}{2}
	\bigg[
	\frac{\delta_\mu^0 Ha}{ \Delta x^2}
	- 
	\partial_\mu
	\bigg(
	\frac{ 1 + \ln\big( \frac{1}{4} H^2 aa' \Delta x^2 \big) }{ \Delta x^2 }
	\bigg)
	\bigg]
	\, ,
\\
&
\frac{1}{\Delta x^{D-2}} \partial_\mu \partial'_\nu \frac{1}{\Delta x^{D-2}}
	\ \overset{D\to4}{\longsim} \ 
	\frac{ \big( D \partial_\mu \partial{}'_\nu - \eta_{\mu\nu} \partial \!\cdot\! \partial' \big) }
		{2(D\!-\!1)}
	\bigg[
	\frac{ \pi^{\frac{D}{2} }  \mu^{D-4} \, i \delta^D(x\!-\!x') }
		{ (D\!-\!3)(D\!-\!4) \, \Gamma\big( \frac{D-2}{2} \big) }
	+
	\frac{ \partial \!\cdot\! \partial'}{8} 
		\bigg(
		\frac{ \ln(\mu^2 \Delta x^2) }{ \Delta x^2 }
		\bigg)
	\bigg]
	\, ,
\\
&
\frac{1}{\Delta x^{D-2}} \partial_\mu \partial'_\nu \frac{1}{\Delta x^{D-4}}
	\overset{D\to4}{\longsim} \ 
	(D\!-\!4)
	\bigg\{
	\eta_{\mu\nu} 
	\bigg[
	\frac{ \pi^{\frac{D}{2} }  \mu^{D-4} \, i \delta^D(x\!-\!x') }
		{ (D\!-\!3) (D\!-\!4) \, \Gamma\big( \frac{D-2}{2} \big) }
	+
	\frac{   \partial \!\cdot\! \partial'}{8}
		\frac{ \ln(\mu^2 \Delta x^2) }{ \Delta x^2 }
	\bigg]
	\!+
	\frac{\partial_\mu \partial{}'_\nu}{4} \frac{1}{\Delta x^2}
	\bigg\}
	 ,
\\
&
\bigg[ \frac{1}{\Delta x^{D-4}} - \Big( \frac{H^2aa'}{4} \Big)^{\!\frac{D-4}{2}} \bigg]
	\partial_\mu \partial'_\nu \frac{1}{\Delta x^{D-2}}
	\ \overset{D\to4}{\longsim} \ 
	(D\!-\!4) \bigg\{
	\eta_{\mu\nu} 
	\bigg[
	\frac{\pi^{\frac{D}{2} }  \mu^{D-4} \, i \delta^D(x\!-\!x') }
		{ (D\!-\!3)(D\!-\!4) \, \Gamma\big( \frac{D-2}{2} \big) }
\\
&	\hspace{1.cm}
	+
	\frac{ \partial \!\cdot\! \partial'}{8} \bigg(
		\frac{ \ln(\mu^2 \Delta x^2) }{ \Delta x^2 }
		\bigg)
	\bigg]
	-
	\frac{\partial_\mu \partial{}'_\nu}{2}
	\bigg(
	\frac{ \frac{3}{2} + \ln\big( \frac{1}{4} H^2 aa' \Delta x^2 \big) }{ \Delta x^2 }
	\bigg)
	+
	\frac{1}{2}
	\big( H a \delta_\mu^0  \partial'_\nu + H a' \delta_\nu^0 \partial_\mu \big)
	\frac{1}{\Delta x^2}
	\bigg\}
	\, .
\nonumber 
\end{align}
\end{subequations}
%

\section{Identities for extracting derivatives and simplifying
expressions}
\label{app: Identities for extracting derivatives}

This appendix collects the identities used in
Secs.~\ref{subsubsec: Contracting tensor structures} and
\ref{subsubsec: Consolidation and renormalization} to
rearrange derivative structures and extract derivatives from
nonlocal terms. Throughout this appendix, the symbol
$\longrightarrow$ denotes equivalence under integration by parts, reflection of derivatives in the self-mass kernel, and, where indicated, the use of other propagator identities. 
We first group the identities used in
Sec.~\ref{subsubsec: Contracting tensor structures} into
three classes.

\subsection{Identities used with tensor contractions}
\label{subapp: Identities used with tensor contractions}

The first class consists of identities that reorganize the
derivative structure of local terms:
\begin{align}
\big( a'^2 \, \overline{\partial} \!\cdot\! \partial
    + a^2 \, \overline{\partial}{}' \!\cdot\! \partial' \big)
    i \delta^D(x\!-\!x')
    \longrightarrow{}&
    2aa' \big(
        \overline{\partial} \!\cdot\! \overline{\partial}{}'
        +
        3 H^2aa'
        \big)
    i \delta^D(x\!-\!x')
    \, ,
\label{LocalId1}
\\
H aa' \big( a' \overline{\partial}{}_0
	+ a \overline{\partial}{}'_0 \big) i \delta^D(x\!-\!x')
	\longrightarrow{}&
    - 3 H^2 (aa')^2 i \delta^D(x\!-\!x')
    \, ,
\label{LocalId2}
\\
H \big( a \partial{}'_0 \!+\! 
    a' \partial{}_0 \big) i \delta^D(x\!-\!x')
	\longrightarrow{}&
    H^2 aa' i \delta^D(x\!-\!x') \, ,
\label{LocalId3}
\\
H (a \!+\! a') \big[ \,
    \overline{\partial}{}_0
    ( \overline{\partial}{}' \!\cdot\! \partial')
    +
    \overline{\partial}{}'_0 
    ( \overline{\partial}{} \!\cdot\! \partial )
    \big]
    i \delta^D(x\!-\!x')
    \longrightarrow{}&
    - 2H^2 aa'
    \overline{\nabla} \!\cdot\! \overline{\nabla}{}'
        i \delta^D(x\!-\!x')
    \, .
\label{LocalId4}
\end{align}
These identities follow from integration by parts.
The second class reorganizes derivatives acting on
nonlocal terms that depend on relative coordinates only:
\begin{align}
\Big[ a'{}^2 \big( \overline{\partial}{} \!\cdot\! \partial 
			\!+\! 2Ha\overline{\partial}{}_0 \big)
		+ a^2 \big( \overline{\partial}{}' \!\cdot\! \partial{}'
			\!+\! 2Ha'\overline{\partial}{}'_0 \big)
		\Big] f(\Delta x^2)
    \longrightarrow{}&
    - \big( a'^4 \overline{\mathcal{D}} + a^4 \overline{\mathcal{D}}{}'
		\big) \frac{f(\Delta x^2)}{(aa')^2}
    \, ,
\label{ReflSymmId1}
\\
H(a\!+\!a') \big[
    \overline{\partial}{}'_0 (\overline{\partial} 
        \!\cdot\! \partial ) 
	+ \overline{\partial}{}_0 ( \overline{\partial}{}' \!\cdot\! \partial' ) \big]
	f(\Delta x^2)
	\longrightarrow{}&
    -
    H^2(a^2 \!+\! a'^2)
    \overline{\nabla} \!\cdot\! \overline{\nabla}{}'
        f(\Delta x^2) 
        \, .
\label{ReflSymmId2}
\end{align}
The third class reduces the derivative order of singular
nonlocal terms. These identities follow from integration by
parts together with Eq.~(\ref{FlatPropEOM}):
\begin{align}
H^2 (aa')^2 \big( 
    a'^2 \overline{\partial}{} \!\cdot\! \partial
        + a^2 \overline{\partial}{}' \!\cdot\! \partial' 
        \big)
	\frac{1}{y}
    \longrightarrow{}&
    - 8 \pi^2 (aa')^2 i \delta^4(x\!-\!x')
    - \frac{4 H^2 (aa')^3}{\Delta x^2}
    \, ,
\label{RedId1}
\\
aa' \big(
    \overline{\partial} \!\cdot\! \partial
        + \overline{\partial}{}' \!\cdot\! \partial' \big)
    \frac{1}{\Delta x^2}
	\longrightarrow{}&
	8 \pi^2 aa' 
    i \delta^4(x\!-\!x')
	-
	H aa'
    \big( a' \partial_0 \!+\! a \partial{}'_0 \big)
		\frac{1}{\Delta x^2}
	\, ,
\label{RedId2}
\\
aa'
(\overline{\partial} \!\cdot\! \partial) (\overline{\partial}{}' \!\cdot\! \partial')
	\frac{1}{\Delta x^2}
	\longrightarrow{}&
	-
	4 \pi^2 aa'
    \overline{\partial} \!\cdot\! \overline{\partial}{}'
	i \delta^4(x\!-\!x')
	+
	H^2 (aa')^2
    \overline{\nabla}\!\cdot\! \overline{\nabla}{}'
	\frac{1}{\Delta x^2}
	\, .
\label{RedId3}
\end{align}
%

\subsection{Identities used with consolidation}
\label{subapp: Identities used with consolidation}

In Sec.~\ref{subsubsec: Consolidation and renormalization}
we use additional identities to extract derivatives from the
finite nonlocal terms appearing in contributions $I$ and
$I\!I$ to the 3-vertex diagram. We list these identities below 
and briefly describe their derivation. Several of them rely on 
the following two identities for extracting derivatives:
\begin{align}
f(y) \partial_\mu g(y)
    ={}&
    \partial_\mu I \Big[ f(y) 
        \frac{\partial g(y)}{\partial y} \Big]
        \, ,
\label{ExtractingSingle}
\\
f(y) \partial_\mu \partial_\nu' g(y)
    ={}&
    \partial_\mu \partial'_\nu
    I^2 \Big[ f(y) \frac{\partial^2 g(y)}{\partial y^2} \Big]
    +
    (\partial_\mu \partial'_\nu y) 
        I\Big[ \frac{\partial f(y)}{\partial y} 
        \frac{\partial g(y)}{\partial y} \Big]
    \, ,
\label{ExtractingDouble}
\end{align}
where the mixed derivative of $y$ is
\begin{equation}
\partial_\mu\partial_\nu' y 
    =
    - H^2 aa' \big(
        2\eta_{\mu\nu}+\delta_\mu^0\delta_\nu^0 y \big)
    + H \big( a\delta_\mu^0\partial_\nu'
        + a'\delta_\nu^0\partial_\mu \big) y
    \, .
\label{yDerivatives}
\end{equation}
We also use the vertex identity
\begin{equation}
a^2 g(x) \, \overline{\partial} \!\cdot\! \partial f(x)
    \longrightarrow
    \frac{1}{2} g(x) \big( \widetilde{\mathcal{D}} 
        - \overline{\mathcal{D}} - \mathcal{D} \big) f(x)
        \, ,
\label{VertexIdentity}
\end{equation}
where $\widetilde{\mathcal{D}}$ is understood to act on the
function $g(x)$.

\paragraph{Contribution $I$.}

For contribution $I$, three identities are
required. The first identity is
\begin{align}
\frac{H^4(aa')^2}{y}
	\big(
	a'{}^2 \, \overline{\partial}{} \!\cdot\! \partial 
	+
	a^2 \, \overline{\partial}{}' \!\cdot\! \partial{}'
	\big) \mathcal{F}_m^1
    \longrightarrow{}&
    -
    H^4 \big( a'^4 \overline{\mathcal{D}}
        +
        a^4 \overline{\mathcal{D}}{}' \big)
        I \Big[ \frac{1}{y} 
            \frac{\partial \mathcal{F}_m^0}{\partial y} \Big]
    -
    \frac{4 (m^2 \!-\! 2H^2) H^2 (aa')^3}{\Delta x^2}
\nonumber \\
&
    -
    8 (m^2\!-\!2H^2) \pi^2 i \delta^4(x\!-\!x')
    \, .
\label{contIidentity1}
\end{align}
To derive it, we first use Eq.~(\ref{ExtractingSingle}) to
extract one derivative. We then express the result in terms
of $\mathcal{F}_m^0$ using Eq.~(\ref{F1toF0}). The remaining 
singular term is then reduced using
\begin{equation}
\big( \mathcal{D} \!-\! 2H^2a^4 \big) \frac{1}{y} = 
    \frac{4\pi^2}{H^2} i \delta^4(x\!-\!x') \, ,
\label{conformalEOM}
\end{equation}
which is the rescaled equation of motion for the
conformally coupled scalar propagator.

The second identity is
\begin{equation}
\frac{H^4(aa')^2}{y}
	\Big[
	( \overline{\partial} \!\cdot\! \partial{}' ) 
    (\overline{\partial}{}' \!\cdot\! \partial)
	-
	( \overline{\partial}{} \!\cdot\! \partial ) 
    ( \overline{\partial}{}' \!\cdot\! \partial{}' )
	\Big]
	\mathcal{F}_m^1
\longrightarrow
    H^5 (aa')^2 ( a \!+\! a' ) 
    \overline{\nabla} \!\cdot\! \overline{\nabla}{}'
	( \overline{\partial}{}'_0 \!+\! \overline{\partial}_0 )
    \bigg(
    \frac{\partial \mathcal{F}_m^0}{\partial y}
    +
    I \Big[ \frac{1}{y} 
        \frac{\partial \mathcal{F}_m^0}{\partial y} \Big]
    \bigg)
    \, .
\label{contIidentity2}
\end{equation}
To derive it, we first decompose the derivative operators on
the left-hand side into temporal and spatial parts:
\begin{equation}
( \overline{\partial} \!\cdot\! \partial{}' ) 
    (\overline{\partial}{}' \!\cdot\! \partial)
	-
	( \overline{\partial}{} \!\cdot\! \partial ) 
    ( \overline{\partial}{}' \!\cdot\! \partial{}' )
    =
    -
    \Big[ \overline{\partial}_0
        (\overline{\nabla}{}' \!\cdot\! \nabla)
        +
        \overline{\partial}{}'_0
        (\overline{\nabla} \!\cdot\! \nabla{}')
        \Big]
    ( \partial_0 \!+\! \partial{}'_0 )
    \, .
\end{equation}
After reflecting the spatial derivatives appropriately, the
purely spatial terms cancel. The remaining temporal
derivatives are simplified using
\begin{equation}
\partial_0 y = Hay - 2H(a \!-\! a' )
\, ,
\qquad 
\partial_0'y = Ha'y + 2H(a \!-\! a' ) \, ,
\qquad 
(\partial_\mu \!+\! \partial_\mu') y = \delta_\mu^0 H(a \!+\! a') y \, .
\label{yIds}
\end{equation}
which follows from the definition of the de Sitter
invariant length function $y$ given in~(\ref{yDef}). 
The remaining inner derivative is then extracted using
Eq.~(\ref{ExtractingSingle}), after which the result can be
rewritten entirely in terms of $\mathcal{F}_m^0$ using
Eq.~(\ref{F1toF0}).

The third identity is
\begin{align}
\MoveEqLeft[4]
\frac{H^4(aa')^2}{y}
	( \overline{\partial} \!\cdot\! \overline{\partial}{}' ) 
    (\partial{}' \!\cdot\! \partial)
	\mathcal{F}_m^1
    \longrightarrow
    (m^2 \!-\! 2H^2)
    aa'
	( \overline{\partial} \!\cdot\! \overline{\partial}{}' ) 
    H \big( a\partial'_0 \!+\! a'\partial_0 \big)
	\frac{1}{\Delta x^2}
\nonumber \\
&
    -
    \frac{m^2H^4(aa')^2 ( a^2 \!+\! a'^2 )}{2}
	( \overline{\partial} \!\cdot\! \overline{\partial}{}' ) 
	\frac{\mathcal{F}_m^0}{y}
    +
    H^5(aa')^2
	( \overline{\partial} \!\cdot\! \overline{\partial}{}' ) 
    \big( a\partial'_0 \!+\! a'\partial_0 \big)
	I \Big[
        \frac{1}{y}\frac{\partial \mathcal{F}_m^0}{\partial y}
        \Big]
\nonumber \\
&
    +
    \frac{H^5(aa')^2(a\!+\!a')}{2}
    ( \overline{\partial} \!\cdot\! \overline{\partial}{}' )
    \Big[
    \overline{\partial}{}'_0 \!+\! \overline{\partial}_0
    \!-\! H(a\!+\!a')
    \Big]
    \frac{\partial\mathcal{F}_m^0}{\partial y}
    \, .
\label{contIidentity3}
\end{align}
We derive it by first decomposing the inner derivative
contraction as
\begin{equation}
\partial \!\cdot\! \partial'
    =
    \frac{1}{2} \big[ (\partial \!+\! \partial')^2
        - \partial^2 - \partial'^2
        \big]
        \, .
\end{equation}
For the first term, Eq.~(\ref{yIds}) eliminates the spatial
derivatives from the symmetric combination
$\partial\!+\!\partial'$. We then act once with the resulting
symmetric time derivative and extract the remaining
derivative using Eq.~(\ref{ExtractingSingle}). For the
remaining two terms, we rewrite the flat-space
d'Alembertians in terms of their de Sitter counterparts,
\begin{equation}
\partial^2 = \frac{1}{a^2} \mathcal{D} + 2Ha\partial_0 \, ,
\qquad\quad
\partial'^2 = \frac{1}{a'^2} \mathcal{D}' + 2Ha'\partial'_0 
\, ,
\end{equation}
and use Eq.~(\ref{eqF1}) to evaluate their action. 
The remaining temporal derivatives are then extracted using 
Eq.~(\ref{ExtractingSingle}) and rearranged with 
Eq.~(\ref{yIds}), yielding Eq.~(\ref{contIidentity3}).

\paragraph{Contribution $I\!I$.}
For contribution $I\!I$, nine additional identities are
required. 
The first two identities are
\begin{align}
\MoveEqLeft[2]
H^2(aa')^2 \big[ C \!+\! \ln(aa') \big] 
    \big( \overline{\partial} \!\cdot\! \partial
        \!+\! m^2 a^2 \!+\!
        \overline{\partial}{}' \!\cdot\! \partial'
        \!+\! m^2 a'^2 \big)
    \Big( \frac{4}{y} \!+\!  \mathcal{F}_m^0 \Big)
    \longrightarrow
    H^3(aa')^2
    \big( a \overline{\partial}_0
        \!+\!
        a' \overline{\partial}{}'_0
        \big)
    \Big( \frac{4}{y} \!+\!  \mathcal{F}_m^0 \Big)
\nonumber \\
&   \hspace{1cm}
    -
    H^2
    \Big[ a'^2
        \big( \overline{\mathcal{D}} \!-\! m^2 a^4 \big)
        + 
        a^2
        \big( \overline{\mathcal{D}}{}' \!-\! m^2 a'^4 \big)
        \Big]
    \big[ C \!+\! \ln(aa') \big] 
    \Big( \frac{4}{y} \!+\!  \mathcal{F}_m^0 \Big)
    \, ,
\label{contIIidentity1}
\\
\MoveEqLeft[2]
H^2(aa')^2 \big[ C \!+\! \ln(aa') \big]
    \big( \overline{\partial} \!\cdot\! \partial
        \!+\! m^2 a^2\big)
    \big( \overline{\partial}{}' \!\cdot\! \partial'
        \!+\! m^2 a'^2 \big)
    \Big( \frac{4}{y} \!+\! \mathcal{F}_m^0 \Big)
\nonumber \\
&
    \longrightarrow
    48 \pi^2 H^2 (aa')^2 i \delta^4(x\!-\!x')
    +
    H^2\big( \overline{\mathcal{D}} \!-\! m^2 a^4 \big)
    \big( \overline{\mathcal{D}}{}' \!-\! m^2 a'^4 \big)
    \big[ C \!+\! \ln(aa') \big]
    \Big( \frac{4}{y} \!+\! \mathcal{F}_m^0 \Big)
    \, .
\label{contIIidentity2}
\end{align}
Both identities follow from integration by parts used to 
extract derivatives. In deriving Eq.~(\ref{contIIidentity2}), 
we additionally use Eq.~(\ref{eqF}) to isolate the local term.

The next pair of identities is
\begin{align}
&
H^2 (aa')^2
    \big( 2 \overline{\partial}{} \!\cdot\! \partial 
        \!+\! m^2 a^2
		\!+\!
        2 \overline{\partial}{}' \!\cdot\! \partial{}' 
        \!+\!
        m^2 a'^2 \big)
	\frac{1}{ y }
    \longrightarrow
    -
    16 \pi^2 aa' i \delta^4(x\!-\!x')
    +
    \frac{(m^2 \!-\! 4H^2) aa' (a^2 \!+\! a'^2)}{ \Delta x^2 }
    \, ,
\label{contIIidentity3}
\\
&
H^2(aa')^2
	\Big[
	\big( \overline{\partial}{} \!\cdot\! \partial 
        \!+\! m^2a^2 \big)
        \overline{\partial}{}' \!\cdot\! \partial{}' 
	+
	\overline{\partial}{} \!\cdot\! \partial 
	\big( \overline{\partial}{}' \!\cdot\! \partial{}' 
        \!+\! m^2 a'{}^2 \big)
	\Big]
	\frac{1}{ y }
    \longrightarrow
    -
    4 ( m^2 \!-\! 2H^2 ) H^4
	\frac{(aa')^4}{ y }
\nonumber \\
&   \hspace{7.cm}
    -
    8\pi^2
    aa'
	\Big[ \overline{\partial}{} \!\cdot\! 
        \overline{\partial}{}'
        \!+\! (m^2 \!-\! 2H^2 )aa' \Big]
	i \delta^4(x\!-\!x')
    \, .
\label{contIIidentity4}
\end{align}
These follow from integration by parts, together with
Eq.~(\ref{conformalEOM}), which is used to identify 
the local term.
They are followed by two direct identities obtained by 
integration by parts to extract the internal derivative 
operators:
\begin{align}
\MoveEqLeft[4]
H^2 (aa')^2
	\big(  \overline{\partial}{} \!\cdot\! \partial 
            \!+\! 
            m^2a^2
			\!+\!
            \overline{\partial}{}' \!\cdot\! \partial{}'
            \!+\! m^2a'^2 \big)
	\frac{ \ln ( y ) }{ y }
    \longrightarrow
    - H^2\Big[
	    a'^2 \big( \overline{\mathcal{D}} \!-\! m^2a^4 \big)
	    +
        a^2 \big( \overline{\mathcal{D}}{}' \!-\! m^2a'^4 \big)
        \Big]
	\frac{ \ln ( y ) }{ y }
    \, ,
\label{contIIidentity5}
\\
\MoveEqLeft[4]
H^2(aa')^2
	\big( \overline{\partial}{} \!\cdot\! \partial \!+\! m^2a^2 \big)
		\big( \overline{\partial}{}' \!\cdot\! \partial{}' \!+\! m^2 a'{}^2 \big)
	\frac{ \ln ( y ) }{ y }
    \longrightarrow
	H^2\big( \overline{\mathcal{D}} \!-\! m^2a^4 \big)
		\big( \overline{\mathcal{D}}{}' \!-\! m^2 a'{}^4 \big)
	\frac{ \ln ( y ) }{ y }
    \, .
\label{contIIidentity6}
\end{align}

The remaining three identities are
\begin{align}
\MoveEqLeft[2]
    H^2 (aa')^2 \ln(y)
    \Big[ (\overline{\nabla} \!\cdot\! \overline{\nabla}{}') 
            (\nabla \!\cdot\! \nabla') 
        - (\overline{\nabla} \!\cdot\! \nabla) 
            (\overline{\nabla}{}' \!\cdot\! \nabla') \Big]
                \mathcal{F}_m^0
    \longrightarrow
    - 4 H^4 (aa')^3
    \overline{\nabla} \!\cdot\! \overline{\nabla}{}'
    I \Big[ \frac{1}{y} 
        \frac{\partial \mathcal{F}_m^0}{\partial y} \Big]
    \, ,
\label{contIIidentity7}
\\
\MoveEqLeft[2]
    H^2 (aa')^2
	\ln(y)
	\Big[
	\overline{\nabla}{} \!\cdot\! \nabla
		\big( \overline{\partial}{}' \!\cdot\! \partial'
            \!+\! m^2a'^2 \big)
    +
    \overline{\nabla}{}' \!\cdot\! \nabla'
		\big( \overline{\partial} \!\cdot\! \partial
            \!+\! m^2a^2 \big)
	\Big]
	\mathcal{F}_m^0
\nonumber \\
&
    \longrightarrow
    - H^2
    \overline{\nabla}{} \!\cdot\! \overline{\nabla}{}'
	\Big[
		a'^2 \big( \overline{\mathcal{D}} \!-\! m^2a^4 \big)
        +
		a^2 \big( \overline{\mathcal{D}}{}' \!-\! m^2a'^4 \big)
	\Big]
    I \Big[ \ln(y) 
    \frac{\partial \mathcal{F}_m^0}{\partial y} \Big]
    -
    4 H^4 (aa')^3
	\overline{\nabla}{} \!\cdot\! \overline{\nabla}{}'
	I \Big[ \frac{1}{y}
    \frac{\partial \mathcal{F}_m^0}{\partial y} \Big]
\nonumber \\
&   \hspace{0.8cm}
    +
    H^2
    \overline{\nabla}{} \!\cdot\! \overline{\nabla}{}'
	\Big[
		a'^2 \big( \overline{\mathcal{D}}
            \!-\! Ha^3 \overline{\partial}_0 \big)
        +
		a^2 \big( \overline{\mathcal{D}}{}'
            \!-\! Ha'^3 \overline{\partial}{}'_0 \big)
	\Big]
    I^2 \Big[ \frac{1}{y}
    \frac{\partial \mathcal{F}_m^0}{\partial y} \Big]
    \, ,
\label{contIIidentity8}
\\
\MoveEqLeft[2]
H^2(aa')^2\ln(y)
    \big( \overline{\partial}{} \!\cdot\! \partial 
        \!+\! m^2a^2 \big)
	\big( \overline{\partial}{}' \!\cdot\! \partial{}' 
        \!+\! m^2 a'{}^2 \big)
	\mathcal{F}_m^0
    \longrightarrow
    8 \pi^2 (m^2 \!-\! 2H^2) (aa')^2 i \delta^4(x\!-\!x')
\nonumber \\
&
    +
    H^2 
    \big( \overline{\mathcal{D}} \!-\! m^2a^4 \big)
	\big( \overline{\mathcal{D}}{}' \!-\! m^2 a'{}^4 \big)
    \bigg(
	\ln(y) \mathcal{F}_m^0
    -
    \frac{3}{2} I \Big[ \frac{\mathcal{F}_m^0}{y} \Big]
    \bigg)
    +
	\frac{4(m^2 \!-\! 2H^2) H^2(aa')^3}{\Delta x^2}
\label{contIIidentity9}
\\
&
    -
    2m^2H^4 (aa')^4 I \Big[ \frac{\mathcal{F}_m^0}{y^2} \Big]
    -
    H^2
    \Big[
    a'^4 \big( \overline{\mathcal{D}} \!-\! m^2 a^4 \big)
    +
    a^4 \big( \overline{\mathcal{D}}{}' \!-\! m^2 a'{}^4 \big)
    \Big]
    \bigg(
	\frac{3m^2}{4} I \Big[ \frac{\mathcal{F}_m^0}{y} \Big]
    \!+\!
    H^2 I \Big[ \frac{\mathcal{F}_m^0}{y^2} \Big]
    \bigg)
    \, .
\nonumber 
\end{align}
The first identity is derived by first reflecting the
appropriate derivatives and then applying
Eq.~(\ref{ExtractingDouble}) to extract the pair of
internal derivatives. The second identity is derived
similarly, with Eq.~(\ref{ExtractingDouble}) applied first
and Eq.~(\ref{ExtractingSingle}) then used to extract the
remaining single derivatives.

The final identity is the most involved. Its derivation
begins by integrating by parts to extract one internal
derivative. Equation~(\ref{ExtractingSingle}) is then used
to extract the remaining single derivative in some terms,
while the vertex identity~(\ref{VertexIdentity}) is applied
to the remaining terms together with Eq.~(\ref{eqF0}) and
\begin{equation}
\mathcal{D} \ln(y) =
    H^2a^4 \Big( \frac{4}{y}\!-\!3 \Big)
    \, .
\end{equation}
A final integration by parts then brings the result to the
form shown in Eq.~(\ref{contIIidentity9}).

For the purpose of demonstrating that no local 
secular contributions to the self-mass appear
apart from the ones generated by the UV renormalization,
the result in~(\ref{contIIidentity9}) can further be 
rewritten by using
\begin{equation}
\big( \mathcal{D} \!-\! m^2a^4 \big) 
    I \Big[ \frac{\mathcal{F}_m^0}{y} \Big]
    =
    4 a^4 \bigg[
    H^2 \bigg(
    \frac{\mathcal{F}_m^0}{y} 
    -
    2 I\Big[ \frac{1}{y} \frac{\partial \mathcal{F}_m^0}{\partial y} \Big]
    \bigg)
    -
    \frac{m^2 \!-\! 2H^2}{y}
    +
    C
    \bigg]
    \, .
\label{LastIdentity}
\end{equation}
This identity is derived by multiplying the equation of 
motion~(\ref{contIIidentity9}) by~$1/y$ and integrating. 
The constant~$C$ can be determined by substituting the 
power-series representation~(\ref{FpowerSeries}) into
this identity.

\section{Additional useful identities}
\label{app: Additional useful identities}

\paragraph{Flat space identity.}
The flat-space identity used at the end of
Sec.~\ref{sec: Self-mass in Minkowski space} is
\begin{equation}
2\partial^2 \mathcal{I}
    \Big[ \frac{f_m^0}{\Delta x^4} \Big]
    =
    -
    ( \partial^2 \!-\! m^2 )
        \frac{f_m^0}{\Delta x^2}
    - \partial^2
       \bigg[ \frac{\ln\big( \frac{1}{4} m^2 \Delta x^2 \big) + 1 + 2\gamma_{\scr \rm E}}{\Delta x^2} \bigg]
    \, .
\label{FlatIdentity}
\end{equation}
To derive this identity, we first separate the most
singular part of $f_m^0$ by writing
\begin{equation}
f_m^0 
    = f_m^1 + \ln\Big( \frac{m^2 \Delta x^2}{4} \Big)
    - 1 + 2 \gamma_{\scr \rm E} \, ,
\end{equation}
which follows directly from the definition
in Eq.~(\ref{FlatDefF}).
After this subtraction, 
\begin{equation}
    2\partial^2 \mathcal{I}
    \Big[ \frac{f_m^1}{\Delta x^4} \Big]
    =
    - ( \partial^2 \!-\! m^2 )
        \frac{f_m^1}{\Delta x^2}
    +
    \frac{m^2 \big[ \ln\!\big( \frac{1}{4} m^2 \Delta x^2 \big)
        - 1 + 2 \gamma_{\scr \rm E} \big] }{\Delta x^2}
    \, ,
\label{AppIntermed}
\end{equation}
the remaining terms are sufficiently regular at coincidence 
so the derivatives generate no additional local
distributions. We may therefore express the spacetime 
derivatives in terms of
derivatives with respect to $\Delta x^2$,
\begin{equation}
\bigg[
    4\bigg(
    \Delta x^2 
        \frac{\partial}{\partial \Delta x^2}
    +
    2 \bigg) \frac{\partial}{\partial \Delta x^2}
    -
    m^2
    \bigg]
    \bigg[ \frac{4}{m^2\Delta x^2}
        + \ln\! \Big( \frac{m^2\Delta x^2}{4} \Big)
        - 1 + 2\gamma_{\scr \rm E}
        + f_m^1 \bigg]
    =
    0
    \, .
\end{equation}
This equation is recognized as the off-coincident equation 
of motion for the massive scalar propagator.

\paragraph{De Sitter space identities.}
The de Sitter space generalization of 
identity~(\ref{FlatIdentity}) is
\begin{align}
2 H^4 \mathcal{D} I \Big[ \frac{\mathcal{F}_m^0}{y^2} \Big]
    ={}&
    - H^4 \big( \mathcal{D} \!-\! m^2a^4 \!+\! 2H^2a^4 \big)
    \frac{\mathcal{F}_m^0}{y}
    -
    \frac{H^4 (m^2 \!-\! 2H^2) a^4}{y}
\nonumber \\
&
    -
    H^2 (m^2 \!-\! 2H^2) \big( \mathcal{D} \!-\! 2H^2a^4 \big)
        \bigg( \frac{\ln(y) \!+\! \Psi_m \!+\! 2}{y} \bigg)
        \, .
\end{align}
It can be verified that this expression reduces to
Eq.~(\ref{FlatIdentity}) in the flat-space limit by using
the limits of the relevant quantities in
Eq.~(\ref{FlatLimits}), together with reflection
invariance.

The identity is derived using the same strategy as in flat
space. We first use~(\ref{F1toF0}) to rewrite the 
expression in terms of
$\mathcal{F}_m^1$, which eliminates the most singular 
terms form the equation,
\begin{equation}
2 H^4 \mathcal{D} I \Big[ \frac{\mathcal{F}_m^1}{y^2} \Big]
    =
- H^4 \big( \mathcal{D} \!-\! m^2a^4 \!+\! 2H^2a^4 \big)
    \frac{\mathcal{F}_m^1}{y}
    +
    H^2(m^2\!-\!2H^2)a^4
    \frac{m^2 \big[ \ln(y) \!+\! \Psi_m \big] \!+\! 3 H^2}{y}
        \, .
\end{equation}
Expressing all derivatives acting on functions of $y$ in
terms of derivatives with respect to $y$, this relation
reduces to
\begin{equation}
\bigg[ (4y\!-\!y^2) \frac{\partial^2}{\partial y^2}
    + 4 (2\!-\!y) \frac{\partial}{\partial y}
    - \frac{m^2}{H^2} \bigg]
    \bigg[ \frac{4}{y} + \frac{m^2\!-\!2H^2}{H^2}
        \big[ \ln(y) \!+\! \Psi_m \big]
        + \mathcal{F}_m^1 \bigg]
        =
        0
        \, ,
\end{equation}
which follows from the equation of motion for the full
massive scalar propagator.

\medskip

Two additional de Sitter identities are useful for
rearranging the final self-mass. The first is
\begin{equation}
(a\!-\!a')^2 \partial\!\cdot\!\partial'
    \bigg(
    \frac{\ln(C^2\Delta x^2)}{\Delta x^2}
    \bigg)
    =
    2Haa'\big( a' \partial_0 \!+\! a \partial'_0 \big)
    \frac{1}{\Delta x^2}
    \, ,
\end{equation}
which is valid for an arbitrary dimensionful
constant~$C$ inside the logarithm. It follows by evaluating 
the derivatives on both sides explicitly. The second identity,
\begin{equation}
H \big( a'\partial_0 + a\partial'_0 + 3Haa' \big) f(y)
    = - \frac{1}{2} \nabla \!\cdot\! \nabla' 
    I \big[ f(y) \big]
    \, ,
\end{equation}
has been reported previously 
in~\cite{Glavan:2020zne,Glavan:2022dwb}.



\begin{thebibliography}{99}

\bibitem{Parker:1968mv}
L.~Parker,
``Particle creation in expanding universes,''
Phys. Rev. Lett. \textbf{21} (1968), 562-564

\bibitem{Starobinsky:1979ty}
A.~A.~Starobinsky,
``Spectrum of relict gravitational radiation and the early state of the universe,''
JETP Lett. \textbf{30} (1979), 682-685

\bibitem{Miao:2006gj}
S.~P.~Miao and R.~P.~Woodard,
``Gravitons Enhance Fermions during Inflation,''
Phys. Rev. D \textbf{74} (2006), 024021
[arXiv:gr-qc/0603135 [gr-qc]].

\bibitem{Glavan:2013jca}
D.~Glavan, S.~P.~Miao, T.~Prokopec and R.~P.~Woodard,
``Electrodynamic Effects of Inflationary Gravitons,''
Class. Quant. Grav. \textbf{31} (2014), 175002
[arXiv:1308.3453 [gr-qc]].

\bibitem{Wang:2014tza}
C.~L.~Wang and R.~P.~Woodard,
``Excitation of Photons by Inflationary Gravitons,''
Phys. Rev. D \textbf{91} (2015) no.12, 124054
[arXiv:1408.1448 [gr-qc]].

\bibitem{Tan:2021lza}
L.~Tan, N.~C.~Tsamis and R.~P.~Woodard,
``How inflationary gravitons affect gravitational radiation,''
Phil. Trans. Roy. Soc. Lond. A \textbf{380} (2021), 0187
[arXiv:2107.13905 [gr-qc]].

\bibitem{Glavan:2021adm}
D.~Glavan, S.~P.~Miao, T.~Prokopec and R.~P.~Woodard,
``Large logarithms from quantum gravitational corrections to a massless, minimally coupled scalar on de Sitter,''
JHEP \textbf{03} (2022), 088
[arXiv:2112.00959 [gr-qc]].

\bibitem{Tan:2022xpn}
L.~Tan, N.~C.~Tsamis and R.~P.~Woodard,
``How Inflationary Gravitons Affect the Force of Gravity,''
Universe \textbf{8} (2022) no.7, 376
[arXiv:2206.11467 [gr-qc]].

\bibitem{Glavan:2016bvp}
D.~Glavan, S.~P.~Miao, T.~Prokopec and R.~P.~Woodard,
``One loop graviton corrections to dynamical photons in de Sitter,''
Class. Quant. Grav. \textbf{34} (2017) no.8, 085002
[arXiv:1609.00386 [gr-qc]].

\bibitem{Kamenshchik:2021tjh}
A.~Y.~Kamenshchik, A.~A.~Starobinsky and T.~Vardanyan,
``Massive scalar field in de Sitter spacetime: a two-loop calculation and a comparison with the stochastic approach,''
Eur. Phys. J. C \textbf{82} (2022) no.4, 345
[arXiv:2109.05625 [gr-qc]].

\bibitem{Arkani-Hamed:2015bza}
N.~Arkani-Hamed and J.~Maldacena,
``Cosmological Collider Physics,''
[arXiv:1503.08043 [hep-th]].

\bibitem{Tsamis:1992xa}
N.~C.~Tsamis and R.~P.~Woodard,
``The Structure of perturbative quantum gravity on a De Sitter background,''
Commun. Math. Phys. \textbf{162} (1994), 217-248

\bibitem{Woodard:2004ut}
R.~P.~Woodard,
``De Sitter breaking in field theory,''
published in {\it Deserfest: A celebration of the life and works of Stanley Deser} (World Scientific, Hackensack, 2006) eds. J. T. Liu, M. J. Duff, K. S. Stelle and R. P. Woodard,
p. 339
[arXiv:gr-qc/0408002 [gr-qc]].

\bibitem{Miao:2011fc}
S.~P.~Miao, N.~C.~Tsamis and R.~P.~Woodard,
``The Graviton Propagator in de Donder Gauge on de Sitter Background,''
J. Math. Phys. \textbf{52} (2011), 122301
[arXiv:1106.0925 [gr-qc]].

\bibitem{Mora:2012zi}
P.~J.~Mora, N.~C.~Tsamis and R.~P.~Woodard,
``Graviton Propagator in a General Invariant Gauge on de Sitter,''
J. Math. Phys. \textbf{53} (2012), 122502
[arXiv:1205.4468 [gr-qc]].

\bibitem{Glavan:2015ura}
D.~Glavan, S.~P.~Miao, T.~Prokopec and R.~P.~Woodard,
``Graviton Loop Corrections to Vacuum Polarization in de Sitter in a General Covariant Gauge,''
Class. Quant. Grav. \textbf{32} (2015) no.19, 195014
[arXiv:1504.00894 [gr-qc]].

\bibitem{Donoghue:1994dn}
J.~F.~Donoghue,
``General relativity as an effective field theory: The leading quantum corrections,''
Phys. Rev. D \textbf{50} (1994), 3874-3888
[arXiv:gr-qc/9405057 [gr-qc]].

\bibitem{Donoghue:1993eb}
J.~F.~Donoghue,
``Leading quantum correction to the Newtonian potential,''
Phys. Rev. Lett. \textbf{72} (1994), 2996-2999
[arXiv:gr-qc/9310024 [gr-qc]].

\bibitem{Bjerrum-Bohr:2002aqa}
N.~E.~J.~Bjerrum-Bohr,
``Leading quantum gravitational corrections to scalar QED,''
Phys. Rev. D \textbf{66} (2002), 084023
[arXiv:hep-th/0206236 [hep-th]].

\bibitem{Bjerrum-Bohr:2002gqz}
N.~E.~J.~Bjerrum-Bohr, J.~F.~Donoghue and B.~R.~Holstein,
``Quantum gravitational corrections to the nonrelativistic scattering potential of two masses,''
Phys. Rev. D \textbf{67} (2003), 084033
[erratum: Phys. Rev. D \textbf{71} (2005), 069903]
[arXiv:hep-th/0211072 [hep-th]].

\bibitem{Miao:2017feh}
S.~P.~Miao, T.~Prokopec and R.~P.~Woodard,
``Deducing Cosmological Observables from the S-matrix,''
Phys. Rev. D \textbf{96} (2017) no.10, 104029
[arXiv:1708.06239 [gr-qc]].

\bibitem{Katuwal:2021thy}
S.~Katuwal and R.~P.~Woodard,
``Gauge independent quantum gravitational corrections to Maxwell{\textquoteright}s equation,''
JHEP \textbf{10} (2021), 029
[arXiv:2107.13341 [gr-qc]].

\bibitem{Glavan:2024elz}
D.~Glavan, S.~P.~Miao, T.~Prokopec and R.~P.~Woodard,
``Gauge independent logarithms from inflationary gravitons,''
JHEP \textbf{03} (2024), 129
[arXiv:2402.05452 [hep-th]].

\bibitem{Glavan:2019msf}
D.~Glavan, S.~P.~Miao, T.~Prokopec and R.~P.~Woodard,
``Graviton Propagator in a 2-Parameter Family of de Sitter Breaking Gauges,''
JHEP \textbf{10} (2019), 096
[arXiv:1908.06064 [gr-qc]].

\bibitem{Glavan:2025azq}
D.~Glavan,
``Graviton propagator in de Sitter space in a simple one-parameter gauge,''
JHEP \textbf{05} (2026), 189
[arXiv:2511.13660 [gr-qc]].

\bibitem{Glavan:2026pug}
D.~Glavan, S.~P.~Miao, T.~Prokopec and R.~P.~Woodard,
``Cancellation of one-parameter graviton gauge dependence in the effective scalar field equation in de Sitter,''
JHEP \textbf{04} (2026), 159
[arXiv:2602.07908 [hep-th]].

\bibitem{Bogoliubov:1957gp}
N.~N.~Bogoliubov and O.~S.~Parasiuk,
``On the Multiplication of the causal function in the quantum theory of fields,''
Acta Math. \textbf{97}, 227-266 (1957)

\bibitem{Hepp:1966eg}
K.~Hepp,
``Proof of the Bogolyubov-Parasiuk theorem on renormalization,''
Commun. Math. Phys. \textbf{2}, 301-326 (1966)

\bibitem{Zimmermann:1968mu}
W.~Zimmermann,
``The power counting theorem for minkowski metric,''
Commun. Math. Phys. \textbf{11}, 1-8 (1968)

\bibitem{Zimmermann:1969jj}
W.~Zimmermann,
``Convergence of Bogolyubov's method of renormalization in momentum space,''
Commun. Math. Phys. \textbf{15}, 208-234 (1969)

\bibitem{Kahya:2007bc}
E.~O.~Kahya and R.~P.~Woodard,
``Quantum Gravity Corrections to the One Loop Scalar Self-Mass during Inflation,''
Phys. Rev. D \textbf{76} (2007), 124005
[arXiv:0709.0536 [gr-qc]].

\bibitem{Glavan:2020gal}
D.~Glavan, S.~P.~Miao, T.~Prokopec and R.~P.~Woodard,
``Single graviton loop contribution to the self-mass of a massless, conformally coupled scalar on a de Sitter background,''
Phys. Rev. D \textbf{101} (2020) no.10, 106016
[arXiv:2003.02549 [gr-qc]].

\bibitem{Grillo:1999ew}
N.~Grillo,
``Scalar matter coupled to quantum gravity in the causal approach: Finite one loop calculations and perturbative gauge invariance,''
Annals Phys. \textbf{287} (2001), 153-190
[arXiv:hep-th/9912128 [hep-th]].

\bibitem{tHooft:1974toh}
G.~'t Hooft and M.~J.~G.~Veltman,
``One-loop divergencies in the theory of gravitation,''
Ann. Inst. H. Poincare Phys. Theor. A \textbf{20} (1974) no.1, 69-94

\bibitem{Capper:1979ej}
D.~M.~Capper,
``A general gauge graviton look calculation,''
J. Phys. A \textbf{13} (1980), 199

\bibitem{Schwinger:1960qe}
J.~S.~Schwinger,
``Brownian motion of a quantum oscillator,''
J. Math. Phys. \textbf{2} (1961), 407-432

\bibitem{Mahanthappa:1962ex}
K.~T.~Mahanthappa,
``Multiple production of photons in quantum electrodynamics,''
Phys. Rev. \textbf{126} (1962), 329-340

\bibitem{Bakshi:1962dv}
P.~M.~Bakshi and K.~T.~Mahanthappa,
``Expectation value formalism in quantum field theory. 1.,''
J. Math. Phys. \textbf{4} (1963), 1-11

\bibitem{Bakshi:1963bn}
P.~M.~Bakshi and K.~T.~Mahanthappa,
``Expectation value formalism in quantum field theory. 2.,''
J. Math. Phys. \textbf{4} (1963), 12-16

\bibitem{Keldysh:1964ud}
L.~V.~Keldysh,
``Diagram Technique for Nonequilibrium Processes,''
Sov. Phys. JETP \textbf{20} (1965), 1018-1026

\bibitem{Chou:1984es}
K.~c.~Chou, Z.~b.~Su, B.~l.~Hao and L.~Yu,
``Equilibrium and Nonequilibrium Formalisms Made Unified,''
Phys. Rept. \textbf{118} (1985), 1-131

\bibitem{Jordan:1986ug}
R.~D.~Jordan,
``Effective Field Equations for Expectation Values,''
Phys. Rev. D \textbf{33} (1986), 444-454

\bibitem{Ford:2004wc}
L.~H.~Ford and R.~P.~Woodard,
``Stress tensor correlators in the Schwinger-Keldysh formalism,''
Class. Quant. Grav. \textbf{22} (2005), 1637-1647
[arXiv:gr-qc/0411003 [gr-qc]].

\bibitem{Chernikov:1968zm}
N.~A.~Chernikov and E.~A.~Tagirov,
``Quantum theory of scalar field in de Sitter space-time,''
Ann. Inst. H. Poincare Phys. Theor. A \textbf{9} (1968) no.2, 109-141

\bibitem{Onemli:2002hr}
V.~K.~Onemli and R.~P.~Woodard,
``Superacceleration from massless, minimally coupled phi**4,''
Class. Quant. Grav. \textbf{19} (2002), 4607
[arXiv:gr-qc/0204065 [gr-qc]].

\bibitem{Glavan:2026bfw}
D.~Glavan, S.~P.~Miao, T.~Prokopec and R.~P.~Woodard,
``Universal Secular External Leg Corrections for Gauge Independent Scalar Self-Mass on de Sitter,''
[arXiv:2609.10365 [gr-qc]].

\bibitem{Burko:2002ge}
L.~M.~Burko, A.~I.~Harte and E.~Poisson,
``Mass loss by a scalar charge in an expanding universe,''
Phys. Rev. D \textbf{65} (2002), 124006
[arXiv:gr-qc/0201020 [gr-qc]].

\bibitem{Akhmedov:2010ah}
E.~T.~Akhmedov, A.~Roura and A.~Sadofyev,
``Classical radiation by free-falling charges in de Sitter spacetime,''
Phys. Rev. D \textbf{82} (2010), 044035
[arXiv:1006.3274 [gr-qc]].

\bibitem{Glavan:2019yfc}
D.~Glavan, S.~P.~Miao, T.~Prokopec and R.~P.~Woodard,
``Breaking of scaling symmetry by massless scalar on de Sitter,''
Phys. Lett. B \textbf{798} (2019), 134944
[arXiv:1908.11113 [gr-qc]].

\bibitem{Glavan:2026jxe}
D.~Glavan and D.~Jaramillo-Garrido,
``On cosmological properties of black-hole hair in linearly coupled scalar-Gauss-Bonnet theory,''
[arXiv:2605.14132 [gr-qc]].

\bibitem{Starobinsky:1986fx}
A.~A.~Starobinsky,
``Stochastic de Sitter (inflationary) stage in the early Universe,''
Lect. Notes Phys. \textbf{246} (1986), 107-126

\bibitem{Starobinsky:1994bd}
A.~A.~Starobinsky and J.~Yokoyama,
``Equilibrium state of a selfinteracting scalar field in the de Sitter background,''
Phys. Rev. D \textbf{50} (1994), 6357-6368
[arXiv:astro-ph/9407016 [astro-ph]].

\bibitem{Miao:2021gic}
S.~P.~Miao, N.~C.~Tsamis and R.~P.~Woodard,
``Summing inflationary logarithms in nonlinear sigma models,''
JHEP \textbf{03} (2022), 069
[arXiv:2110.08715 [gr-qc]].

\bibitem{Woodard:2023rqo}
R.~P.~Woodard and B.~Yesilyurt,
``Unfinished business in a nonlinear sigma model on de Sitter background,''
JHEP \textbf{06} (2023), 206
[arXiv:2302.11528 [gr-qc]].

\bibitem{Glavan:2023lvw}
D.~Glavan and T.~Prokopec,
``When tadpoles matter: one-loop corrections for spectator Higgs in inflation,''
JHEP \textbf{10} (2023), 063
[arXiv:2306.11162 [hep-ph]].

\bibitem{Litos:2023nvj}
C.~Litos, R.~P.~Woodard and B.~Yesilyurt,
``Large inflationary logarithms in a nontrivial nonlinear sigma model,''
Phys. Rev. D \textbf{108} (2023) no.6, 065001
[arXiv:2306.15486 [gr-qc]].

\bibitem{Glavan:2023tet}
D.~Glavan, S.~P.~Miao, T.~Prokopec and R.~P.~Woodard,
``Explaining large electromagnetic logarithms from loops of inflationary gravitons,''
JHEP \textbf{08} (2023), 195
[arXiv:2307.09386 [gr-qc]].

\bibitem{Miao:2024nsz}
S.~P.~Miao, N.~C.~Tsamis and R.~P.~Woodard,
``Summing gravitational effects from loops of inflationary scalars,''
Class. Quant. Grav. \textbf{41} (2024) no.21, 215007
[arXiv:2405.01024 [gr-qc]].

\bibitem{Foraci:2024vng}
A.~J.~Foraci and R.~P.~Woodard,
``Resumming photon loops for inflationary gravity,''
Phys. Rev. D \textbf{111} (2025) no.6, 063531
[arXiv:2412.11022 [gr-qc]].

\bibitem{Foraci:2024cwi}
A.~J.~Foraci and R.~P.~Woodard,
``Resumming fermion loops for inflationary gravity,''
Phys. Lett. B \textbf{868} (2025), 139649
[arXiv:2501.01972 [gr-qc]].

\bibitem{Glavan:2020zne}
D.~Glavan, A.~Marunovi{\'c}, T.~Prokopec and Z.~Zahraee,
``Abelian Higgs model in power-law inflation: the propagators in the unitary gauge,''
JHEP \textbf{09} (2020), 165
[arXiv:2005.05435 [gr-qc]].

\bibitem{Glavan:2022dwb}
D.~Glavan and T.~Prokopec,
``Photon propagator in de Sitter space in the general covariant gauge,''
JHEP \textbf{05} (2023), 126
[arXiv:2212.13982 [gr-qc]].


\end{thebibliography}
\end{document}